\documentclass[twocolumn]{aastex62}

\newcommand{\pkg}[1]{\texttt{#1}}

\shorttitle{Performance of the Gemini Planet Imager Non-Redundant Mask}
\shortauthors{Greenbaum et al.}

\begin{document}

\title{Performance of the Gemini Planet Imager Non-Redundant Mask and spectroscopy of two close-separation binaries HR 2690 and HD 142527}

\author[0000-0002-7162-8036]{Alexandra Z. Greenbaum}
\affiliation{Department of Astronomy, University of Michigan, Ann Arbor, MI 48109, USA}

\correspondingauthor{Alexandra Z. Greenbaum}
\email{azgreenb@umich.edu}

\author{Anthony Cheetham}
\affiliation{D\'epartement d'Astronomie, Universit\'e de Gen\`eve, 51 chemin des Maillettes, 1290 Versoix, Switzerland }

\author[0000-0003-1251-4124]{Anand Sivaramakrishnan}
\affiliation{Space Telescope Science Institute, Baltimore, MD 21218, USA}

\author[0000-0002-9667-2244]{Fredrik T. Rantakyr\"o}
\affiliation{Gemini Observatory, Casilla 603, La Serena, Chile}

\author[0000-0002-5092-6464]{Gaspard Duch\^ene}
\affiliation{Department of Astronomy, University of California, Berkeley, CA 94720, USA}
\affiliation{Univ. Grenoble Alpes/CNRS, IPAG, F-38000 Grenoble, France}

\author{Peter Tuthill}
\affiliation{Sydney Institute for Astronomy, School of Physics, The University of Sydney, NSW 2006, Australia}

\author[0000-0002-4918-0247]{Robert J. De Rosa}
\affiliation{Department of Astronomy, University of California, Berkeley, CA 94720, USA}
\affiliation{Kavli Institute for Particle Astrophysics and Cosmology, Stanford University, Stanford, CA 94305, USA}

\author[0000-0001-7130-7681]{Rebecca Oppenheimer}
\affiliation{Department of Astrophysics, American Museum of Natural History, New York, NY 10024, USA}

\author[0000-0003-1212-7538]{Bruce Macintosh}
\affiliation{Kavli Institute for Particle Astrophysics and Cosmology, Stanford University, Stanford, CA 94305, USA}

\author{S. Mark Ammons}
\affiliation{Lawrence Livermore National Laboratory, 7000 East Ave., Livermore, CA 94550}

\author[0000-0002-5407-2806]{Vanessa P. Bailey}
\affiliation{Jet Propulsion Laboratory, California Institute of Technology, Pasadena, CA 91109, USA}

\author[0000-0002-7129-3002]{Travis Barman}
\affiliation{Lunar and Planetary Laboratory, University of Arizona, Tucson AZ 85721, USA}

\author{Joanna Bulger}
\affiliation{Subaru Telescope, NAOJ, 650 North A{'o}hoku Place, Hilo, HI 96720, USA}

\author[0000-0003-2871-4390]{Andrew Cardwell}
\affiliation{Large Binocular Telescope Observatory, University of Arizona, 933 N. Cherry Ave, Room 552, Tucson, AZ 85721, USA}

\author[0000-0001-6305-7272]{Jeffrey Chilcote}
\affiliation{Department of Physics, University of Notre Dame, 225 Nieuwland Science Hall, Notre Dame, IN, 46556, USA}

\author[0000-0003-0156-3019]{Tara Cotten}
\affiliation{Department of Physics and Astronomy, University of Georgia, Athens, GA 30602, USA}

\author{Rene Doyon}
\affiliation{Institut de Recherche sur les Exoplan{\`e}tes, D{\'e}partement de Physique, Universit{\'e} de Montr{\'e}al, Montr{\'e}al QC, H3C 3J7, Canada}

\author[0000-0002-0176-8973]{Michael P. Fitzgerald}
\affiliation{Department of Physics \& Astronomy, University of California, Los Angeles, CA 90095, USA}

\author[0000-0002-7821-0695]{Katherine B. Follette}
\affiliation{Physics and Astronomy Department, Amherst College, 21 Merrill Science Drive, Amherst, MA 01002, USA}

\author[0000-0003-3978-9195]{Benjamin L. Gerard}
\affiliation{University of Victoria, Department of Physics and Astronomy, 3800 Finnerty Rd, Victoria, BC V8P 5C2, Canada} 
\affiliation{National Research Council of Canada Herzberg, 5071 West Saanich Rd, Victoria, BC, V9E 2E7, Canada}

\author[0000-0002-4144-5116]{Stephen J. Goodsell}
\affiliation{Gemini Observatory, 670 N. A'ohoku Place, Hilo, HI 96720, USA}

\author{James R. Graham}
\affiliation{Department of Astronomy, University of California, Berkeley, CA 94720, USA}

\author[0000-0003-3726-5494]{Pascale Hibon}
\affiliation{Gemini Observatory, Casilla 603, La Serena, Chile}

\author[0000-0003-1498-6088]{Li-Wei Hung}
\affiliation{Department of Physics \& Astronomy, University of California, Los Angeles, CA 90095, USA}

\author{Patrick Ingraham}
\affiliation{Large Synoptic Survey Telescope, 950N Cherry Ave., Tucson, AZ 85719, USA}

\author{Paul Kalas}
\affiliation{Department of Astronomy, University of California, Berkeley, CA 94720, USA}
\affiliation{SETI Institute, Carl Sagan Center, 189 Bernardo Ave.,  Mountain View CA 94043, USA}

\author[0000-0002-9936-6285]{Quinn Konopacky}
\affiliation{Center for Astrophysics and Space Science, University of California San Diego, La Jolla, CA 92093, USA}

\author{James E. Larkin}
\affiliation{Department of Physics \& Astronomy, University of California, Los Angeles, CA 90095, USA}

\author{J\'er\^ome Maire}
\affiliation{Center for Astrophysics and Space Science, University of California San Diego, La Jolla, CA 92093, USA}

\author[0000-0001-7016-7277]{Franck Marchis}
\affiliation{SETI Institute, Carl Sagan Center, 189 Bernardo Ave.,  Mountain View CA 94043, USA}

\author[0000-0002-5251-2943]{Mark S. Marley}
\affiliation{NASA Ames Research Center, Mountain View, CA 94035, USA}

\author[0000-0002-4164-4182]{Christian Marois}
\affiliation{National Research Council of Canada Herzberg, 5071 West Saanich Rd, Victoria, BC, V9E 2E7, Canada}
\affiliation{University of Victoria, 3800 Finnerty Rd, Victoria, BC, V8P 5C2, Canada}

\author[0000-0003-3050-8203]{Stanimir Metchev}
\affiliation{Department of Physics and Astronomy, Centre for Planetary Science and Exploration, The University of Western Ontario, London, ON N6A 3K7, Canada}
\affiliation{Department of Physics and Astronomy, Stony Brook University, Stony Brook, NY 11794-3800, USA}

\author[0000-0001-6205-9233]{Maxwell A. Millar-Blanchaer}
\affiliation{Jet Propulsion Laboratory, California Institute of Technology, Pasadena, CA 91109, USA}
\affiliation{NASA Hubble Fellow}

\author[0000-0002-1384-0063]{Katie M. Morzinski}
\affiliation{Steward Observatory, 933 N. Cherry Ave., University of Arizona, Tucson, AZ 85721, USA}

\author[0000-0001-6975-9056]{Eric L. Nielsen}
\affiliation{SETI Institute, Carl Sagan Center, 189 Bernardo Ave.,  Mountain View CA 94043, USA}
\affiliation{Kavli Institute for Particle Astrophysics and Cosmology, Stanford University, Stanford, CA 94305, USA}

\author{David Palmer}
\affiliation{Lawrence Livermore National Laboratory, Livermore, CA 94551, USA}

\author{Jennifer Patience}
\affiliation{School of Earth and Space Exploration, Arizona State University, PO Box 871404, Tempe, AZ 85287, USA}

\author[0000-0002-3191-8151]{Marshall Perrin}
\affiliation{Space Telescope Science Institute, Baltimore, MD 21218, USA}

\author{Lisa Poyneer}
\affiliation{Lawrence Livermore National Laboratory, Livermore, CA 94551, USA}

\author{Laurent Pueyo}
\affiliation{Space Telescope Science Institute, Baltimore, MD 21218, USA}

\author[0000-0002-9246-5467]{Abhijith Rajan}
\affiliation{Space Telescope Science Institute, Baltimore, MD 21218, USA}

\author[0000-0003-0029-0258]{Julien Rameau}
\affiliation{Institut de Recherche sur les Exoplan{\`e}tes, D{\'e}partement de Physique, Universit{\'e} de Montr{\'e}al, Montr{\'e}al QC, H3C 3J7, Canada}

\author{Naru Sadakuni}
\affiliation{Stratospheric Observatory for Infrared Astronomy, Universities Space Research Association, NASA/Armstrong Flight Research Center, 2825 East Avenue P, Palmdale, CA 93550, USA}

\author[0000-0002-8711-7206]{Dmitry Savransky}
\affiliation{Sibley School of Mechanical and Aerospace Engineering, Cornell University, Ithaca, NY 14853, USA}

\author{Adam C. Schneider}
\affiliation{School of Earth and Space Exploration, Arizona State University, PO Box 871404, Tempe, AZ 85287, USA}

\author[0000-0002-5815-7372]{Inseok Song}
\affiliation{Department of Physics and Astronomy, University of Georgia, Athens, GA 30602, USA}

\author[0000-0003-2753-2819]{Remi Soummer}
\affiliation{Space Telescope Science Institute, Baltimore, MD 21218, USA}

\author{Sandrine Thomas}
\affiliation{Large Synoptic Survey Telescope, 950N Cherry Ave., Tucson, AZ 85719, USA}

\author{J. Kent Wallace}
\affiliation{Jet Propulsion Laboratory, California Institute of Technology, Pasadena, CA 91109, USA}

\author[0000-0003-0774-6502]{Jason J. Wang}
\affiliation{Department of Astronomy, University of California, Berkeley, CA 94720, USA}

\author[0000-0002-4479-8291]{Kimberly Ward-Duong}
\affiliation{Physics and Astronomy Department, Amherst College, 21 Merrill Science Drive, Amherst, MA 01002, USA}

\author[0000-0003-4483-5037]{Sloane Wiktorowicz}
\affiliation{The Aerospace Corporation, 2310 E. El Segundo Blvd., El Segundo, CA 90245}

\author[0000-0002-9977-8255]{Schuyler Wolff}
\affiliation{Leiden Observatory, Leiden University, P.O. Box 9513, 2300 RA Leiden, The Netherlands}

\begin{abstract} 
The Gemini Planet Imager (GPI) contains a 10-hole non-redundant mask (NRM),
enabling interferometric resolution in complement to its coronagraphic
capabilities. The NRM operates both in spectroscopic (integral field
spectrograph, henceforth IFS) and polarimetric configurations.  NRM
observations were taken between 2013 and 2016 to characterize its performance.
Most observations were taken in spectroscopic mode with the goal of obtaining
precise astrometry and spectroscopy of faint companions to bright stars. We
find a clear correlation between residual wavefront error measured by the AO
system and the contrast sensitivity by comparing phase errors in observations
of the same source, taken on different dates. We find a
typical  5-$\sigma$ contrast sensitivity of $2-3~\times~10^{-3}$ at $\sim\lambda/D$.  
We explore the accuracy
of spectral extraction of secondary components of binary systems by recovering
the signal from a simulated source injected into several datasets.
We outline data reduction procedures unique to GPI's IFS and describe a newly
public data pipeline used for the presented analyses. We demonstrate recovery
of astrometry and spectroscopy of two known companions to HR 2690 and HD
142527.  NRM+polarimetry observations achieve differential
visibility precision of $\sigma\sim0.4\%$ in the best case.  We discuss its
limitations on Gemini-S/GPI for resolving inner regions of protoplanetary disks
and prospects for future upgrades. We summarize lessons learned in observing with NRM in
spectroscopic and polarimetric modes. 
\end{abstract}


\section{Introduction \label{sec:intro}}

Exoplanet imaging survey instruments reach deep contrast performance by
attenuating the stellar PSF using a
coronagraph~\citep[e.g.][]{oppenheimer2012,macintosh2014,beuzit2008,liu2010}.
Many designs have significantly reduced sensitivity within a $5 ~\lambda/D $
angular region around the host star, where $\lambda$ is the
wavelength and $D$ is the telescope diameter. High resolution, non-occulting
methods, like non-redundant mask (NRM)
interferometry~\citep[e.g.,][]{baldwin1986,tuthill2000}, complement high
contrast methods by probing small spatial scales at moderate contrast. NRM
coupled with adaptive optics can reach contrast of about 6 magnitudes at
$\lambda/B$, with reduced contrast below $\lambda/B$~\citep{lacour2011}. In
this case $B$ is the longest baseline spanned by the mask (typically close to
the telescope diameter). This complementary high resolution approach can reveal
the presence of close-in structures to bright point sources, particularly
exciting for young protoplanetary systems.  The NRM is especially suited for
multiplicity studies at $<2\lambda/D$ scales. Combined with polarimetry,
resolved polarized structures can be resolved close in to the host star. 

High resolution imaging can play an important role in bridging the gap between
companion point source detection methods. Very high contrast methods probe the
outer architectures of solar systems and have little or no overlap with
astrometry or radial velocity detection sensitivities (the latter in part due
to differences in age sensitivities between RV and imaging). High resolution
methods like NRM are sensitive to a objects at intermediate separations,
especially for sources over $100$pc away. NRM on large ground-based telescopes
has been used to resolve structure in the gaps of transitional
disks~\citep[e.g.,][]{kraus2012, biller2012, sallum2015},
has helped push multiplicity studies to closer
separations~\citep[e.g.][]{kraus2008, sana2014, duchene2018} and track the
orbits of close binaries in combination with radial velocity to determine
dynamical masses for young stellar binaries~\citep{rizzuto2016}.
NRM has also been used for image reconstruction of massive
stars~\citep[e.g.,][]{tuthill1999, norris2012} and
disks~\citep[e.g.,][]{cheetham2015,sallum2015tcha}.

Combining aperture masking with an adaptive optics (AO) system provides stable
observations to take advantage of both image quality provided by AO and
self-calibrating observables measured with a non-redundant pupil.  By splitting
the pupil into a set of unique hole-to-hole baseline pairs, fringe phases and
amplitudes can be measured uniquely, where each fringe is
formed from a pupil baseline. In addition, phase closure calibrates hole-based
phase errors that arise from atmospheric fluctuations and instrument non-common
path aberrations~\citep{jennison1958}.  In the case of extreme adaptive optics
(ex-AO), which uses thousands of actuators to correct small corrugations in the
wavefront, interference fringes are stable over many seconds of integration
making fainter sources more accessible through this method.

In this paper we present results from observations with the Gemini Planet
Imager NRM and discuss the performance and post-processing in detail. Our
analysis provides a comparison with aperture masking on other ground-based
instruments, and demonstrate complementarity with upcoming space-based NRM on
JWST-NIRISS~\citep{doyon2012}. We confirm results using two different data
pipelines detailing the data reduction procedures.  With the release of this
article, we make our primary pipeline public, along with examples of analyses
in this paper.

\section{Implementing NRM on The Gemini Planet Imager}

\subsection{The Gemini Planet Imager's non-redundant mask}
\begin{figure}
    \centering
    \includegraphics[height=1.5in]{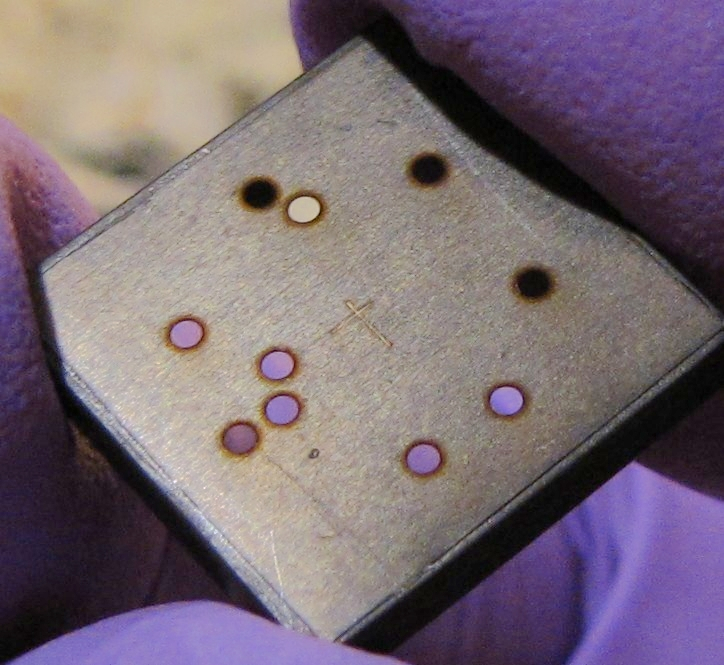}
    \includegraphics[width=1.5in]{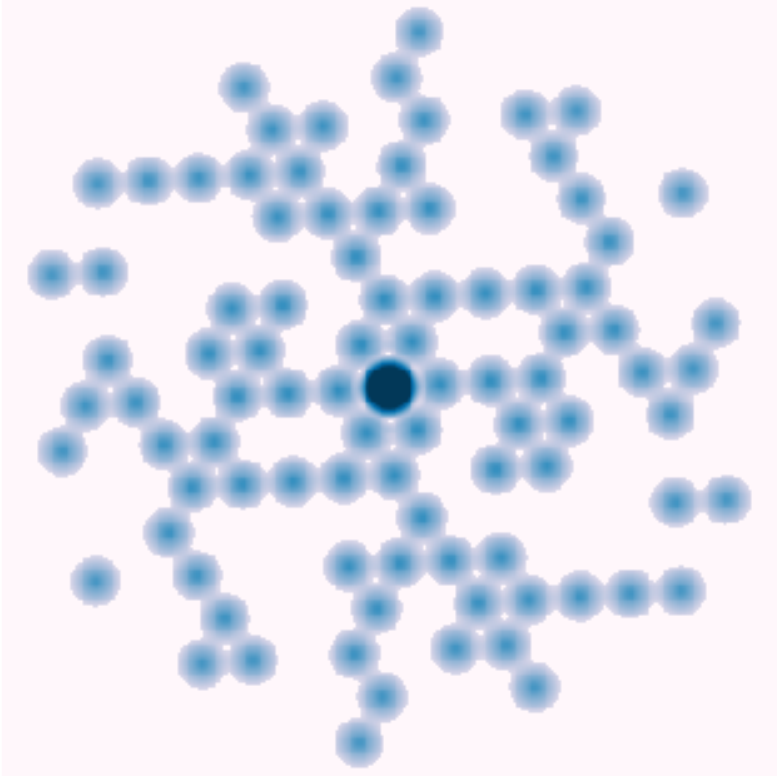}
    \caption{\textbf{Left:} The 10-hole non-redundant mask on the Gemini Planet
Imager. \textbf{Right:} Associated spatial frequency coverage, where the
longest baseline is $6.68~$m.}
    \label{fig:g10s40}
\end{figure}

GPI has a 10-hole non-redundant mask (Fig. \ref{fig:g10s40}) in its apodizer
wheel, a warm pupil located after the deformable mirror. We provide the mask
hole coordinates with respect to the primary mirror in Table
\ref{tab:maskgeom}, including the outer diameter physical size in the apodizer
wheel where the mask sits. This pupil mask transmits roughly 6.2\% of the light
compared to a completely unocculted pupil (not considering spiders, secondary
obstructions, or Lyot stops). The mask forms 45 unique baselines (spatial
frequencies), which correspond to 45 fringes in the image plane. $\lambda/B$
spans $\sim45-330$ mas in H band. There are 120 total combinations of hole
triplets that form closing triangles, and a set of 36 unique triangles that
don't repeat any baseline. 

\begin{table}[htbp!]
\caption{Mask hole dimensions measured in mm from center. \label{tab:maskgeom}}
\begin{center}
\begin{tabular}{c|c}
    X	&	Y  \\ \hline
    -1.061&   -3.882 \\
    -0.389&   -5.192 \\
     2.814&   -0.243 \\
     3.616&    0.995 \\
    -4.419&    2.676 \\
    -1.342&    5.077 \\
    -4.672&   -2.421 \\
     4.157&   -2.864 \\
     5.091&    0.920 \\
     1.599&    4.929 \\
\end{tabular}
\end{center}
Hole diameter: 0.920 mm \\Gemini S outer diameter (OD): 7.770 m (after
baffling) \\Apodizer outer diameter in this re-imaged pupil plane: 11.68 mm.
(Lenox Laser, Glen Arm, MD).\\
Projection of in-pupil coordinates are magnified by a factor of $\sim650$ onto the primary.\\
\end{table} 

GPI’s focal plane masks are implemented as mirrors that reflect the off-axis
light to the science channel and pass the on-axis starlight through a central
hole. In NRM mode, we use a mirror with no hole, so the full field of view
passes to the IFS. However, in coronagraph mode the central starlight is sent
to a tip/tilt sensor for additional low-order correction. Therefore, all
\textit{non-coronagraphic} observations do not benefit from this additional
tip/tilt correction. Small jitter in the image leads to slight smearing of
fringes and reduced contrast. This is worsened in poorer weather conditions,
including high winds. We discuss this in detail in Section \ref{sec:contrast}.

The NRM pupil position for GPI has been measured and fixed to lie entirely
within the pupil and not overlap with any defective actuators or spider
supports. The in-pupil mask coordinates are listed in Table \ref{tab:maskgeom}
and are converted to projected coordinates on the primary mirror by the 
factor between the pupil and primary outer diameter (OD): $7770.1/11.998$\footnote{
future calibration may change this magnification slightly.}.
 The position should not need to be adjusted but any vignetting can be
investigated with the pupil-viewing camera. A detailed discussion of the
procedure to determine the mask orientation and adjusting its position can be
found in \cite{greenbaum2014spie}. Baseline coordinates are computed
as $U_{i,j} = X_i - X_j$,$V_{i,j} = Y_i - Y_j$ for $[i,j]$ combinations, 
where $X$ and $Y$ are the mask hole position in the pupil (Table \ref{tab:maskgeom}).
In the coordinate system used in this work, to reach the detector
orientation the mask coordinates were rotated clockwise by
$114.7^{\circ}$. In Python, Converting the initial baseline vectors $[U_0, V_0]$ into 
vectors rotated by $\theta$, $[U_r, V_r]$ consists of the operation:
$U_{r} = U_0\cos(\theta) - V_0\sin(\theta)$, $V_{r} = U_0\sin(\theta) +
V_0\cos(\theta)$. 

\subsection{Observing Sequences and Calibration}

Uncorrected wavefront and non-common path errors lead to residual phase errors.
At least one nearby calibration source, close in time, (single, unresolved)
should be observed in a sequence. In spectroscopic mode it is less important to
choose a calibration source that matches the target color because the
individual wavelength slices are close to monochromatic. A calibration source
should aim to match the target brightness in 
the wavefront sensing filter (approximately I band). Multiple calibration sources
in a survey-like program can provide a good estimate of systematic calibration
errors~\citep[e.g.,][]{kraus2008}, as long as the sources are observed
consecutively, in similar conditions. However, in the case of observing
individual science targets it may not be practical or efficient to obtain many
calibration sources.

At current operation, it takes approximately 10 minutes to slew to and acquire
a new target. This makes back and forth switching between target and
calibration source time consuming. We have adopted the strategy of observing
the target in full sequence followed by one or two calibration sources.  A
polarimetric sequence additionally involves looping through four half waveplate
angles (HWPAs) per ``integration." While this increases the total integration
on source compared to the spectroscopic mode, polarimetric images are broadband
so each integration is generally shorter. 

Choosing the exposure time for a single integration is a balance between
observation efficiency and minimizing fringe smearing. Typically, we aim for an
exposure time that provides at minimum $3000$ counts in the peak of the raw
detector image and at maximum $14000$ counts to avoid saturation. The total
number of photons collected should satisfy the desired contrast sensitivity. We
discuss systematics that degrade contrast sensitivity beyond photon noise in
Section \ref{sec:contrast}. 

Table \ref{tab:blims} lists the approximate maximum brightness for NRM
observations in each filter combination. All brightness limits and estimated
exposure times approximate and derived empirically from commissioning
observations. 
An empirically-determined exposure time calculator is available in the
\pkg{ImPlaneIA} pipeline\footnote{https://github.com/agreenbaum/ImPlaneIA}.

\begin{table}[htbp!]
\caption{Gemini Planet Imager approximate maximum brightness limits for all NRM
settings. All values are in apparent magnitude in the specified band.}
\label{tab:blims}
\begin{center}
\begin{tabular}{l|c|c|c|c|c}
    MODE	&	Y  & J & H & K1 & K2  \\ \hline\hline
    Spectroscopic & 1.8 & 2.2 & 1.8 & 1.8 & 1.8  \\
    Polarimetric & 3.0 & 3.0 & 3.0 & 3.0 & 3.0
\end{tabular}
\end{center}
\end{table} 

\section{Observations and data reduction}

All observations discussed in this paper were taken on the Gemini Planet Imager
with its 10-hole non-redundant mask, as a part of program GS-ENG-GPI-COM. A
summary of the observations, all taken in stationary pupil mode, is contained
in Table \ref{tab:obs}. The observations presented in this paper focus mostly
on point sources in a range of conditions to determine contrast limits and
polarization precision, as well as two binary systems at different contrast
ratios.

\begin{table*}[htbp]
\caption{Summary of observations presented in this paper, indicating date
string, source name, observing mode, total integration time, and various
observatory parameters. All observations are taken in stationary pupil mode so
that the sky rotates with respect to the detector. \label{tab:obs}}
\begin{center}
\begin{tabular}{l|c|c|c|c|c|c|c|c|c|c}
\hline
\hline %\tabularnewline
Date & Source & Mode & Single& $N_{exp}$ & $=$ Total\footnote{Single
integration $\times$ Number of exposures = Total integration} &
Seeing\footnote{DIMM (Differential Image Motion Monitor)}  &
WFE\footnote{Residual WFE (wavefront error) measured from GPI's AO system.} &
Airmass & Wind\footnote{Ground-layer wind measurement} & sky rot
\tabularnewline
YYMMDD&  & & [s] &  & [s]& ($\arcsec$) & [nm] &  & [m/s] & [deg]
\tabularnewline
\hline
%\tabularnewline
131211 & HR 2690 & NRM Spect - H & 59.6 & 8 & 476.8& 0.67 & 116.15 & 1.23 &
0.49 & 0.021 \tabularnewline
& HR 2716 & NRM Spect - H & 59.6 & 8 & 476.8 & 0.58 & 121.80 & 1.23 & 0.36 &
0.37 \tabularnewline
& HR 2839 & NRM Spect - H & 43.6 & 8 & 348.8 & 0.51 & 134.2 & 1.23 & 0.40 &
0.234 \tabularnewline
140324 & HD 63852 & NRM Spect - H & 1.5 & 20 & 30.0 & 0.87 & 81.99 & 1.17 &
0.41 & 0.67 \tabularnewline
140511 & HD 63852 & NRM Spect - H & 1.5 & 20 & 30.0 & 0.81 & 160.94 & 1.55 &
11.5 & 0.5 \tabularnewline
& Internal & NRM Spect - H & 1.5 & 63 & 94.5 & N/A & 32.82 & N/A & N/A & N/A
\tabularnewline
140512 & HD 142527 & NRM Spect - J & 59.6 & 9 & 536.4 & 1.4 & 190.57 & 1.03 &
8.5 & 11.4 \tabularnewline
& HD 142695 & NRM Spect - J & 53.8 & 8 & 430.4 & 1.4 & 177.77 & 1.04 & 8.6 &
5.0 \tabularnewline
160504& HIP 74604  & NRM pol - K1 & 4.4 & 40 & 176.0 & 2.19 & 144.23 & 1.08 &
4.6 & 1.5 \tabularnewline
\end{tabular}
\end{center}
\end{table*}

During commissioning in December of 2013 we observed the known binary HR 2690
($\Delta H\sim2$) and two unresolved calibration sources HR 2716 and HR 2839.
This sequence of observations was chosen to demonstrate the recovery of a
moderate contrast binary system for proof of concept. In March of 2014 we
observed bright single source HD 63852 to estimate contrast limits compared to
the ideal case of the internal source. In May of 2014 we returned to this
source, providing a comparison between observing epochs. We also observed HD
142527, which contains an M-dwarf companion, HD 142527 B ($\Delta J \sim4.6$)
to demonstrate deeper contrast retrieval of a known binary companion. For this
dataset we observed two calibration sources HD 142695 and HD 142384, though the
latter was found to be a close binary after our
observations~\citep{lebouquin2014}. Details of the analysis are in \S
\ref{sec:spect}. In May of 2016 we took polarimetric observations of bright
unresolved sources to determine calibration limit and assess systematic biases.
We present one example, HIP 74604, our best dataset, and discuss polarimetric
sensitivity in \S \ref{sec:pol}.

\subsection{Raw data reduction \label{sec:drp}}

The data are processed from raw 2D detector exposures into datacubes of images
at each wavelength or polarization through the GPI Data Reduction Pipeline
(DRP)~\citep{drp}. Wavelength calibration is performed with Argon arc lamp
exposures. Shifts in the location of the spectra due to flexure are calibrated
by arc lamp exposures taken close in time to each set of observations
\cite{wolff2014}. For polarimatry data we use the recipe template for
polarization data taken with the NRM called ``Basic NRM Polarization Datacube
Extraction," which performs the polarimetric spot calibration, smooths
polarization calibration, subtracts a dark background, corrects for 2D flexure,
removes microphonics noise, and interpolates bad pixels in the raw frame before
assembling the polarization data cube. Details of DRP primitives can be found
in online documentation\footnote{http://docs.planetimager.org/pipeline/}.

\subsection{Extracting Fringe observables}
\label{sec:datareduction}
We measure fringe phases and amplitudes from reduced datacubes using two
different aperture masking pipelines, the Sydney University pipeline, based in
IDL, and a pipeline implementing the Lacour-Greenbaum (LG)
algorithm~\citep{greenbaum2015}, based in Python. The former analyzes images in
the Fourier domain. The latter measures fringes in the image plane. 

The Fourier plane approach used in the Sydney pipeline measures the phases and
square-visibilities directly from the Fourier transform of the image. First,
images are multiplied by a super-Gaussian window function of the form
$e^{-kx^4}$, which has the effect of smoothing in the Fourier plane. Then,
images are Fourier transformed, which separates the information from different
baselines into distinct regions. The phases and visibilities are measured for
all points in a 3-Fourier sampling element radius around the
predicted frequency for each baseline. To calculate the square-visibilities and
phases for each baseline, these measurements are combined by weighting with a
matched filter.
Closure phases are formed by considering sets of 3 baselines that form a
closing triangle (i.e. the vector sum of their frequencies is zero). Rather
than use the weighted phases for each baseline, instead a number of
measurements are calculated from each set of 3 pixels (within a small area
around the predicted frequency of each baseline) that forms a closing triangle.
These are then combined by weighting with a matched
filter~\citep[e.g.][]{monnier1999}. This matched filter approach relies on
pre-computing the expected Fourier-plane profile of NRM images using fixed
values for the size of the pupil mask holes, plate scale and wavelength for
each IFS channel.

The image plane pipeline assumes a plate scale and monochromatic wavelength
(spectroscopic mode) or defined bandpass (polarimetric/broadband mode) and fits
$A' \sin ( k\cdot \Delta x_{i,j}) + B' \cos(k \cdot \Delta x_{i,j}) $ to each
fringe generated by particular hole-pair baselines, where
$A' = A\sin(\Delta\phi)$ and $B' = B\cos(\Delta\phi)$, $\phi$ is the
fringe phase shift, and $\sqrt{A^2 + B^2}$ is the fringe amplitude.
Here, $k=(u,v)$ is the 2D coordinate in the image plane. This
algorithm is described in detail in \S3 of \cite{greenbaum2015}. The sub-pixel
centering of the image is measured by computing $x$ and $y$ tilt in the
numerical Fourier transform of the image. This centroid is used to sample the
model onto oversampled detector pixels, which are then binned to the detector
scale. For NRM+polarimetry (or broadband) images, for which there is dispersion
in the PSF we use filter transmission files available in the GPI DRP and an
approximate source spectrum to model the dispersion.

We compared the two pipelines and confirmed that they yield consistent results.
We show results from the image-plane pipeline in this paper. An image-plane
pipeline using the LG algorithm, \pkg{ImPlaneIA}~\citep{greenbaum2018soft,
pipeline_2018} is available
publicly\footnote{https://github.com/agreenbaum/ImPlaneIA} with further
documentation and examples.

\subsection{Calibration and analysis of fringe observables
\label{sec:observables}}
Both the Sydney and LG pipelines use similar analysis tools following
calculation of fringe observables to produce the results shown in this paper.

For spectroscopic data we compute an average closure phase and standard error
over the set of integrations for each baseline (each mask hole pair for each
wavelength slice). This produces $N_{\mathrm{triangles}}\times N_{\lambda}$ observables.
In this case $N_{\mathrm{triangles}}=120$ in one datacube slice, and $N_{\lambda}=37$.
In general, we do not see a large amount of field rotation in our observation
sequences (see Table \ref{tab:obs}) so we compute an average position and
consider an average parallactic angle. For our observations of HD 142527, which
contains $\sim11^\circ$ of rotation, we compared the results when accounting
for sky rotation by splitting exposures into smaller groups (see \S
\ref{sec:spect} for more details).  We subtract measured average
closure phases from the calibration source(s) from our science target closure
phases and add errors in quadrature. 

Binary detection and contrast limits rely on a model for the fringe visibility
of a binary point source:
\begin{eqnarray}
V_{u,v} = \frac{1+ r e^{-2\pi i(\alpha\cdot u + \delta \cdot v)}}{1+r}
\label{eqn:visibilities}
\end{eqnarray}
where $r$ is the contrast ratio between the secondary and primary, $u,v$ are
the baseline coordinates a given hole pair, and $\alpha,\delta$ are the sky
coordinates of the secondary relative to the primary. The absolute orientation
is calibrated in the standard way for GPI data, accounting for the orientation
of the lenslet array ($+23.5^\circ$), detector, and instrument position angle
(PA).  Plate scale and PA calibrations have been performed by the observation
of astrometric calibrators yielding a pixel scale of $14.166\pm0.007$ and a
north offset of $-0.1^\circ \pm0.13$ \citep{konopacky2014, derosa2015}. The
derotation angle in degrees to place North up is $\mathrm{AVPARANG} -
\mathrm{AVCASSANG} +23.4$. AVPARANG and AVCASSANG are header keywords in GPI
data files. 

In practice closure phase errors are often underestimated from the data,
especially when only one or two calibration sources are observed and systematic
errors cannot be properly determined. We scale the errors by a factor
$\sqrt{N_{holes}/3}$ to account for redundancy from repeating baselines.
Additionally, we add additional constant error to the closure phases so that
the reduced $\chi^2$ is close to 1. 

The binary detection limits reported in this paper are estimated from the
calibrated closure phase errors based on a signal-to-noise ratio (SNR)
threshold, where
\begin{eqnarray} \label{eqn:snr}
SNR &=& \sqrt{\sum_{i=1}^{N_{CP}}CP_{i,\alpha,\delta,r}^2/\sigma_{i,CP}^2}
\end{eqnarray}
Model closure phases are calculated from equation \ref{eqn:visibilities}. Model
phases scale roughly linearly with contrast ratio $r$. We estimate contrast
ratio detection limit at SNR=5 as:
\begin{eqnarray} \label{eqn:r}
&& r_5 = \frac{5\times r_{model}}{SNR}
\end{eqnarray}
To generate contrast curves we compute $r_5$ over a range of separations and
position angles. Sensitivity varies somewhat with position angle based on mask
geometry. GPI's mask has fairly uniform visibility coverage, improved further
in spectroscopic mode by the wavelength axis. 

In polarimetric mode, the light is split with a Wollaston prism into two
orthogonal polarizations. A half-wave plate optic is used to rotate the angle
of polarization during observation~\citep{perrin2015}. This enables a
differential measurement between orthogonal polarizations for both fringe
amplitude and fringe phase. We  compute differential visibilities and
differential closure phases following \cite{norris2015}.

With 4 half-wave plate (HWP) rotations at 0, 22.5, 45, and 67.5 degrees, we can
build up two layers of calibration. First we calibrate orthogonal polarizations
in a single image:
\begin{eqnarray}
CP_{ortho-diff} &=& CP_{channel 1} - CP_{channel 2} \nonumber \\
V_{ortho-diff} &=& \frac{V_{channel 1}}{V_{channel 2}}
\label{eqn:diffchan}
\end{eqnarray}

Next we calibrate orthogonal HWP rotations, for example, HWPA=$0^o$ with
HWPA=$45^o$:
\begin{eqnarray}
CP_{0-45} &=& CP_{diff-0} - CP_{diff-45} \nonumber \\
V_{0-45} &=& \frac{V_{diff-0}}{V_{diff-45}}
\label{eqn:diffang}
\end{eqnarray}
This should remove instrumental effects, which would contribute to all
polarization states.

\section{Spectroscopic mode \& binary contrast performance}
\label{sec:spect}

The spectroscopic mode on GPI provides a nearly monochromatic image at a set of
wavelengths across each filter. Lack of bandwidth smearing makes fringe
extraction straightforward in this configuration. The extra wavelength
dimension provides many more baselines for a single observation,
$N_{\lambda}\times N_{baselines}$ compared to $N_{baselines}$, where
$N_{\lambda}$ is typically 37 for GPI. 

\begin{figure}
\centering
    \includegraphics[width=3.2in]{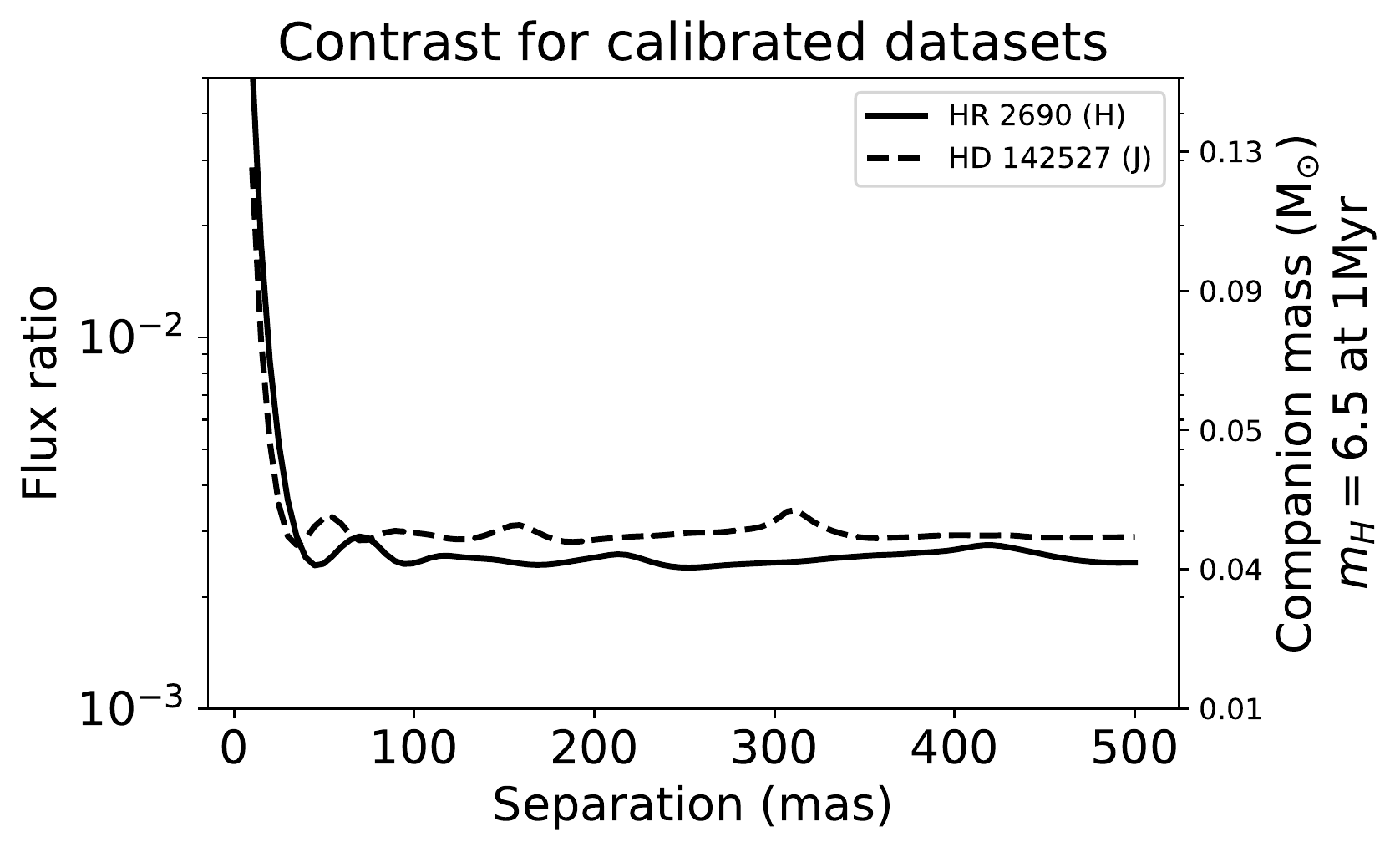}
    \caption{Contrast limits at SNR=5 for the two spectroscopic mode calibrated
datasets analyzed in this paper, for HR 2690 ($\sim8$ minutes in H band) and HD
142527 ($\sim9$ minutes in J band). The righthand vertical axis
shows the corresponding companion mass for given an apparent H magnitude of 6.5
for the primary assuming an age of 1Myr at 140pc using the AMES-Cond models
\citep{baraffe2003}.}
    \label{fig:typicalcontrast}
\end{figure}

\cite{zimmerman2012} demonstrated improved contrast from the set of IFS+NRM
images compared to the combined dataset using the P1640 IFS. We find similar
results when we analyze phase errors measured over all wavelength channels of
the full datacube compared to data collapsed over the wavelength axis. For the
collapsed data we model the PSF as polychromatic considering the approximate
H-band filter throughput profile for GPI. The rest of the analysis is identical
to the typical GPI case described in \S\ref{sec:datareduction}.

\begin{figure*}
\centering
    \includegraphics[width=5in]{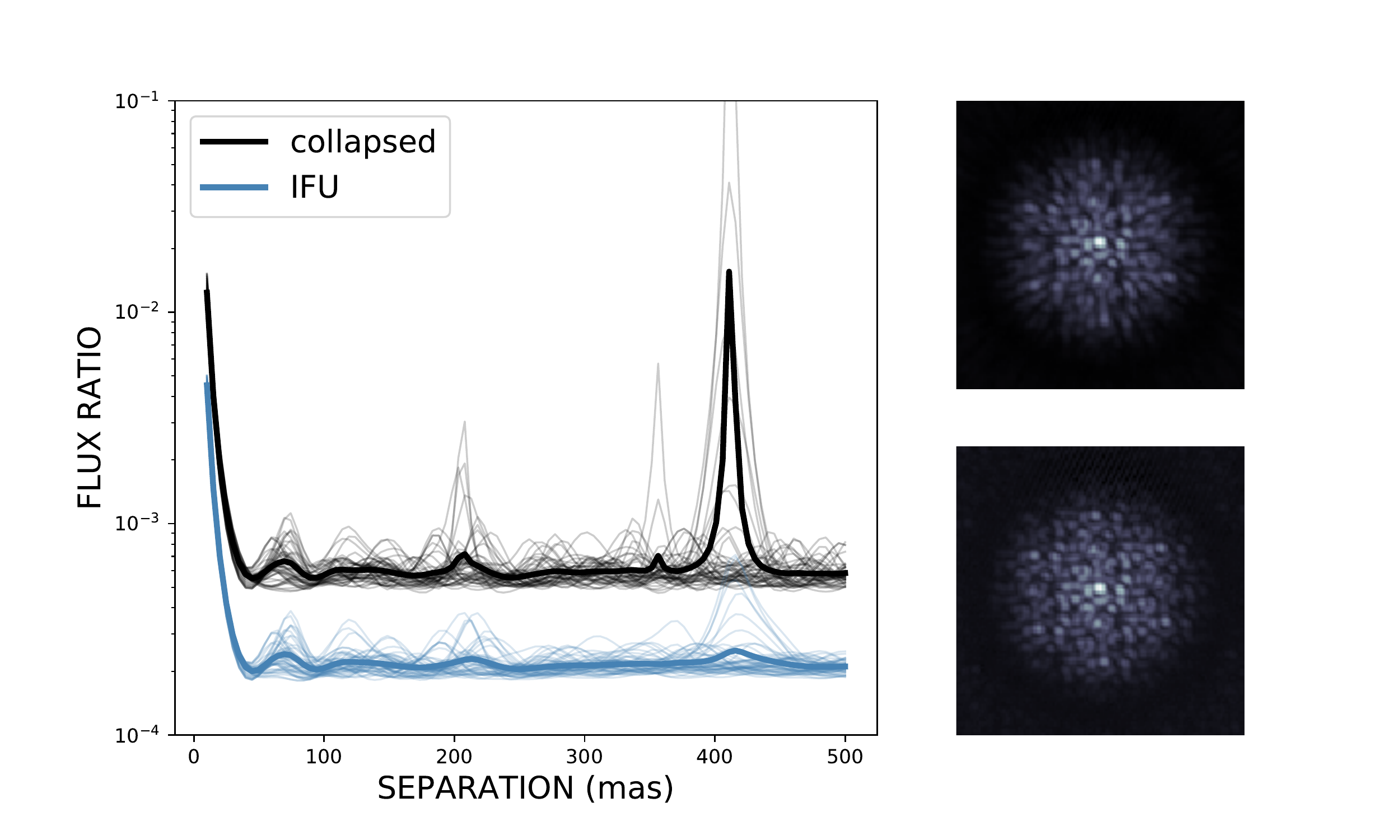}
    \caption{Performance comparison based on internal source data for
$\lambda$-collapsed (black) and IFS data (blue). The contrast is estimated at
SNR=5. The first half of the data are calibrated with the second half,
overestimating performance. The right panel shows snapshots of the data. The IFS
provides twice the sensitivity and smooths out low-sensitivity windows at 200
and 400 mas.}
    \label{fig:ifu}
\end{figure*}

In Figure \ref{fig:ifu} we show an estimated contrast curve for an example
dataset taken with the GPI internal source in the light blue curves, which uses
all wavelength channels. The contrast curve is computed according to Equations
\ref{eqn:snr} and \ref{eqn:r} after scaling the errors by the baseline
redundancy. We also scale the errors by a factor $\sqrt{37/17}$, which roughly
accounts for the fact that we measure $37$ wavelength channels interpolated
over about 17 pixels. The full set of datacubes are split into two halves of
exposures and calibrated against each other. This likely overestimates the
sensitivty, but we consider the relative performance between data taken in
different observing conditions. 
When the data is summed into one polychromatic image, contrast sensitivity is a
factor of $\sim2-3$ worse. The spectroscopic mode is ideal for detection of
faint companions to bright host stars, providing increased signal to noise
overall. The additional spatial frequency coverage reduces regions of very low
sensitivity that arise from the baseline configuration (i.e. the peak of the
collapsed cube curve at $\sim200$ and $\sim400$ mas).

\subsection{Analyzing IFS Data - Simulation Example}
\label{sec:spect_procedure}
Spectral mode datasets can provide robust binary detection, constraining a
companion's position at multiple wavelengths.  We explore errors and biases on
parameter estimation with simulated data of a binary source. The data are
simulated from shifting and adding point source images measured from GPI's
internal light source. Using internal source data ensures there is no resolved
structure in the primary and also that the data still represent aspects of the
GPI PSF that are not modeled (e.g., vibrations, detector effects).
In general, this example will underestimate typical errors for
two reasons: the bright internal source PSF is much more stable and the
secondary companion is simulated from the same data as the PSF calibrator (as
though one had a ``perfect" calibrator). We use this as an example to
demonstrate the approach and provide more practical examples in
\S\ref{sec:hr2690} and \S\ref{sec:142527}. The simulated faint companion 45.5
mas away ($\sim1.2\lambda/D$, $\sim1.0\lambda/B_{max}$) at a position angle of
18.4$^\circ$. We simulate an example flux ratio spectrum between two Phoenix
models~\citep[e.g.,][]{allard2003} at $\mathrm{T}=3240~\mathrm{K}$ and
$\mathrm{T}=5363~\mathrm{K}$ at 10Myr. 
We measure the flux ratio spectrum in the following steps:
\begin{enumerate}
    \item Fit for average flux ratio and common position over all
$N_{\lambda}\times N_{CP}$ observables by MCMC.
    \item Find the flux ratio that minimizes $\chi^2_{binary}$ at the fixed
position determined by the median position parameters recovered in Step 1.
    \item Applying the result from Steps 1 and 2 as a starting guess, use MCMC
to fit a common position and $N_\lambda$ flux ratios (for each wavelength
channel) -- a total of $N_{\lambda}+2$ parameters. 
\end{enumerate}

\textbf{Fit for average flux ratio and common position:} 
We first fit for three parameters in the binary model: position angle,
separation and average contrast using observables from all wavelength channels
using \pkg{emcee}~\citep{dfm2013, dfm2013soft}.  Our posteriors
are localized around the solution, however error between our
simulated parameters and the recovered ones are larger than 1-sigma, indicating
that errors may be underestimated. 

\textbf{Generate an initial estimate for flux ratio spectrum}: Next we fix the
median position and fit for the contrast that minimizes $\chi^2_{binary}$ in
each wavelength channel. This will provide a good starting guess for a finer
fit of the spectrum and position. While it may not be essential to do this
step, it is relatively fast to compute and can be a useful diagnostic before
running a full MCMC fit for all parameters.  Flux ratio errors in each channel
are calculated by including all points on the $\chi^2$ grid where $\chi^2 <
1+\chi^2_{min}$. This is similar to the procedure in  \cite{gauchet2016} for
computing detection maps. However, instead of computing reduced $\chi^2$, we
find that using raw $\chi^2$ with errors scaled by a factor
$\sqrt{N_{holes}/3}$ to account for baseline redundancy, produces fractional
errors consistent with the fractional true error, defined as: $$f_{true} =
\frac{s_{simulated} - s_{recovered}}{s_{simulated}}$$ where $s_{simulated}$ and
$s_{recovered}$ are the simulated and recovered spectra in contrast,
respectively.  This method provides a good estimate of the spectrum across the
band for a moderate contrast binary and is relatively quick to compute, but
does not take into account the position parameter errors.

\textbf{Simultaneous fitting of spectrum and relative astrometry:} Finally, we
fit for the flux ratio in each wavelength channel and common position of the
companion using \pkg{emcee}.  We apply a long burn-in of 5000 iterations with
150 walkers, and run the fit for an additional 5000 iterations.  After an
initial run, we add closure phase error in quadrature to the closure phases
errors so that the reduced $\chi^2$ is roughly equal to 1, in this case
$0.1^\circ$ of additional error. We then recompute this full step. 

\begin{figure}
    \centering
    \includegraphics[width=3.5in]{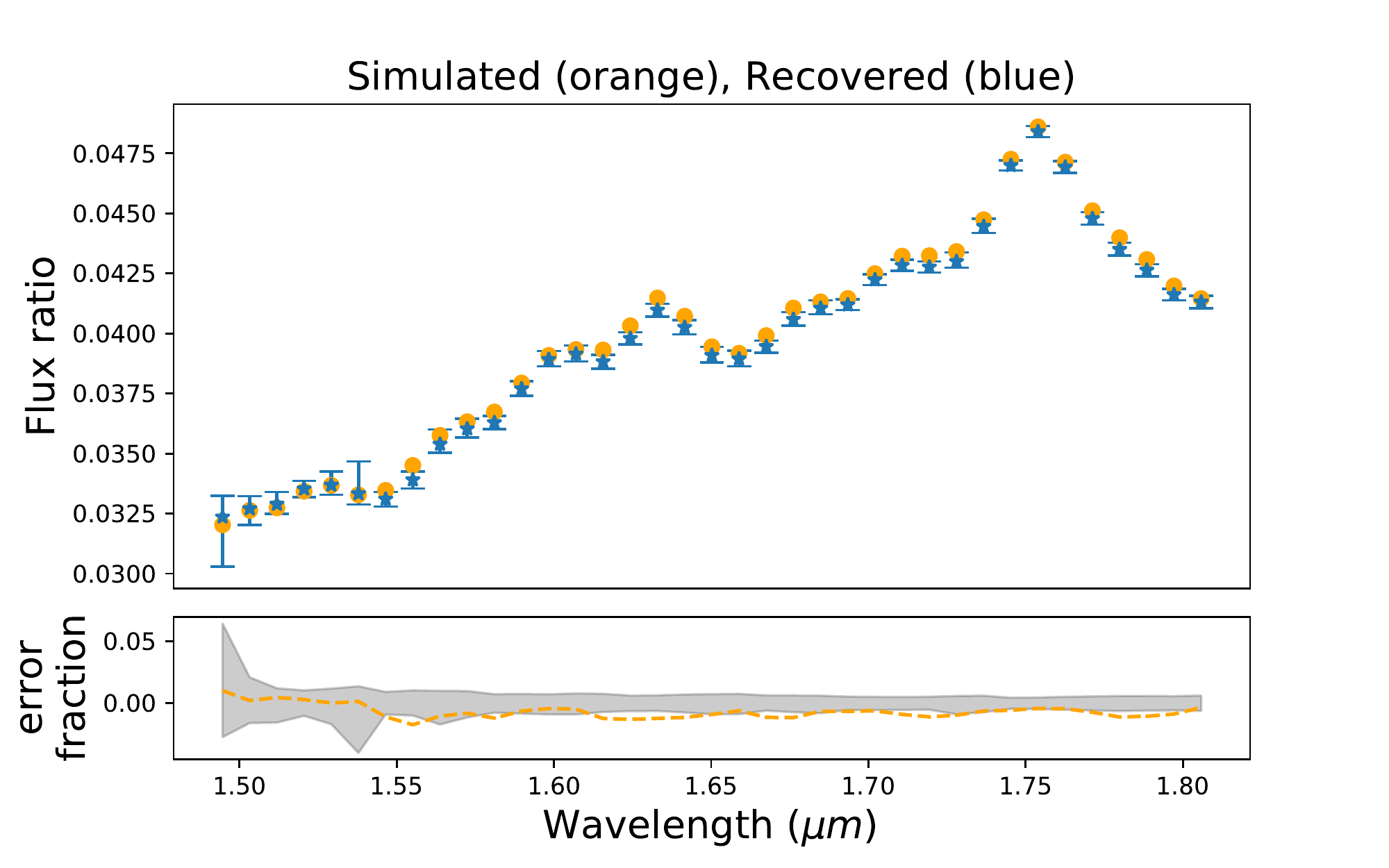}
    \caption{The resulting spectrum measured by MCMC fit over 39 parameters
(flux ratio in 37 wavelength channels, separation, and position angle). The
orange dots represent the simulated spectrum and the blue stars represent the
spectrum recovered by this method with $1\sigma$ errors. The
dashed orange line in the bottom panel shows the fractional error between the
simulated and recovered spectrum, while the gray region shows the fractional
error bounds.
    \label{fig:sim_spectrum_recovery_mcmc}}
\end{figure}

We summarize the results of this procedure in Table
\ref{tab:simastrometry} and Figure \ref{fig:sim_spectrum_recovery_mcmc}.  In
this case, the astrometry changes slightly between the two fits and the true
error is larger than the computed errorbars (which are significantly lower than
for expected on-sky observations that are properly calibrated).  For the
recovered spectrum the contrast in each channel is correct within the
errorbars, with a small bias towards lower flux.

\begin{table} \label{tab:simastrometry}
\caption{Summary of input parameters and results from initial 3-parameter fit,
and the full fit of astrometry and all wavelength channels simultaneously.}
\begin{center}
\begin{tabular}{|l|c|c|c|}
\hline
& Separation & PA & Avg. Contrast \tabularnewline
\hline
Input & $45.4 \mathrm{mas} $ & $18.4^\circ $ & 0.3975 \tabularnewline
3-param & $45.47 \pm 0.03$ & $18.33 \pm 0.02$ & $0.0406 \pm 0.0001$
\tabularnewline
Full fit & $45.24 \pm 0.03$ & $18.36 \pm 0.03$ & $0.0395 \pm 0.003$\footnote{
the average contrast error is computed by adding the error in each channel in
quadrature. This is an overestimation given covariance between
frames.} \tabularnewline
\hline
\end{tabular}
\end{center}
\end{table}

\subsection{Spectral Channel Correlations}
\begin{figure}
    \centering
    \includegraphics[width=3.4in]{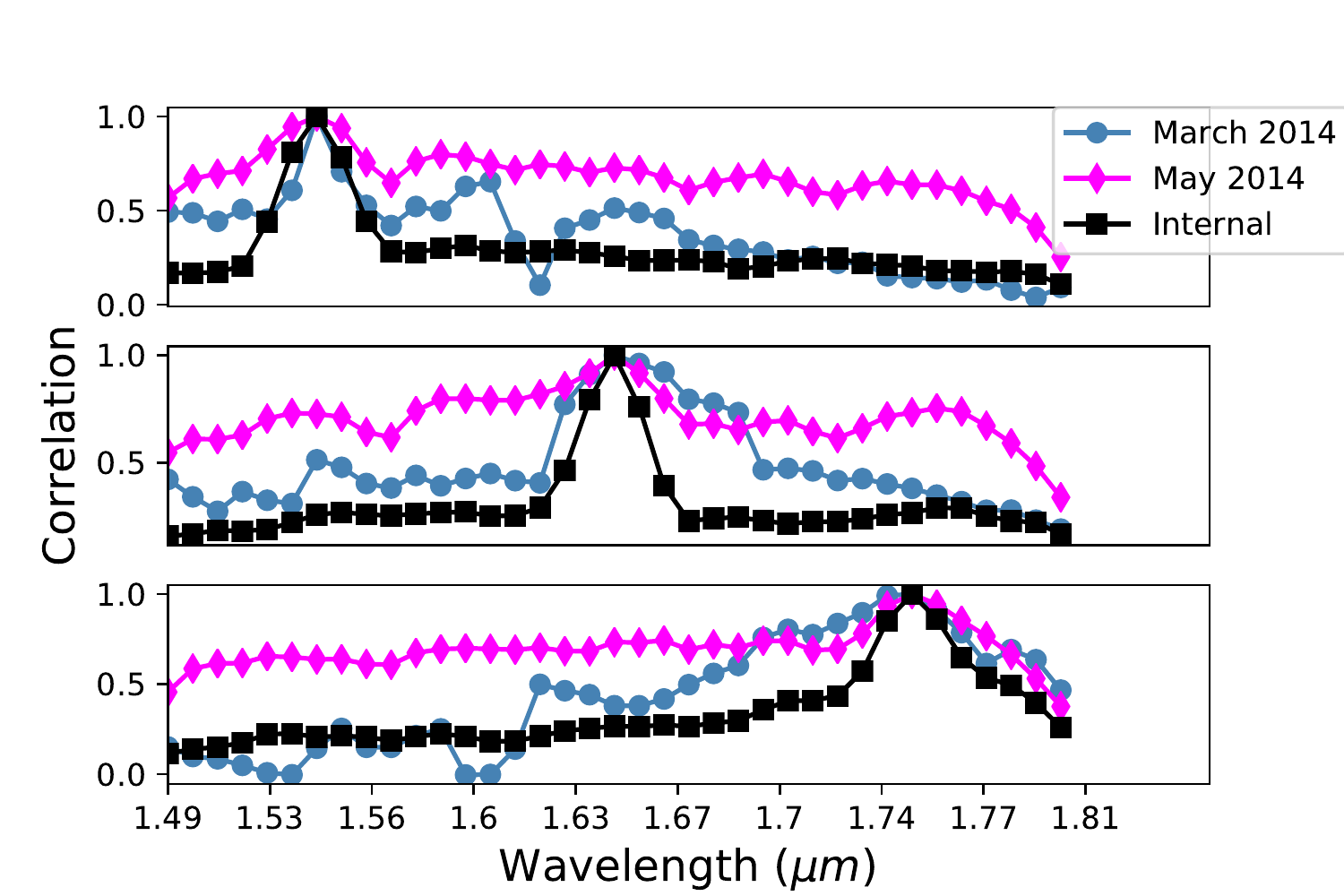}
    \caption{Phase correlations over spectral channel with respect to channels
6 (top), 18 (middle), and 30 (bottom), The internal source data (black squares)
shows low levels of correlation except in the nearest neighboring channels. On
sky images in March 2014 (blue circles), which saw better conditions, and in
May 2014 (pink diamonds), which saw worse conditions, show larger correlation
between channels.}    
    \label{fig:simcorr}
\end{figure}
Following \cite{zimmerman2012} we can describe the correlation of closure
phases between spectral channels. 
The average correlation is defined as:
\begin{eqnarray} 
C(q,w_1;q,w_2) &=& \frac{\langle(\Psi_{q,w1} -
\overline{\Psi}_{q,w1})(\Psi_{q,w2} -
\overline{\Psi}_{q,w2})\rangle}{\sigma_{\Psi_{q,w1}}\sigma_{\Psi_{q,w2}}}
\label{eqn;corr}\\
\overline{C}(w_1;w_2) &=& \sum_{q=0}^{N_{CP -
1}}\frac{Corr(q,w_1;q,w_2)}{N_{CP}} \label{eqn:avgcorr}
\end{eqnarray}
Where $\Psi_{q,w_{i}}$ represents all the measured closure phases of the $q$th
triplet at channel $w_{i}$, $\overline{\Psi}$ is the mean, and $\sigma$ is the
standard deviation.

\cite{zimmerman2012} showed large correlations between spectral channels across
the band for P1640~\citep{oppenheimer2012} NRM IFS images. Some correlation is
expected due to interpolation along the wavelength axis. The simulated dataset,
generated from internal source data, does not suffer from atmospheric
fluctuations. In this case we see a small amount of correlation between
channels except for the nearest neighboring 2-3 channels (Figure
\ref{fig:simcorr}). This is likely dominated by the interpolation. The internal
source data provide an estimate the limiting performance of the instrument. 

For on-sky data, depending on observing conditions we find higher levels of
correlation between spectral channels, beyond the effect of interpolating the
wavelength solution. In Figure \ref{fig:simcorr} we also compare spectral
channel correlations of the two on-sky datasets. In poor conditions (which also
correspond to worse contrast sensitivity) we see a high amount of correlation
across almost all spectral channels. This is likely the result of smearing of
fringes due to vibration and/or non-static phase errors. We further discuss the
differences between these data in Section \ref{sec:spect}.

\subsection{GPI+NRM single source contrast performance \label{sec:contrast}}

\begin{figure*}
\centering
    \includegraphics[scale=0.4]{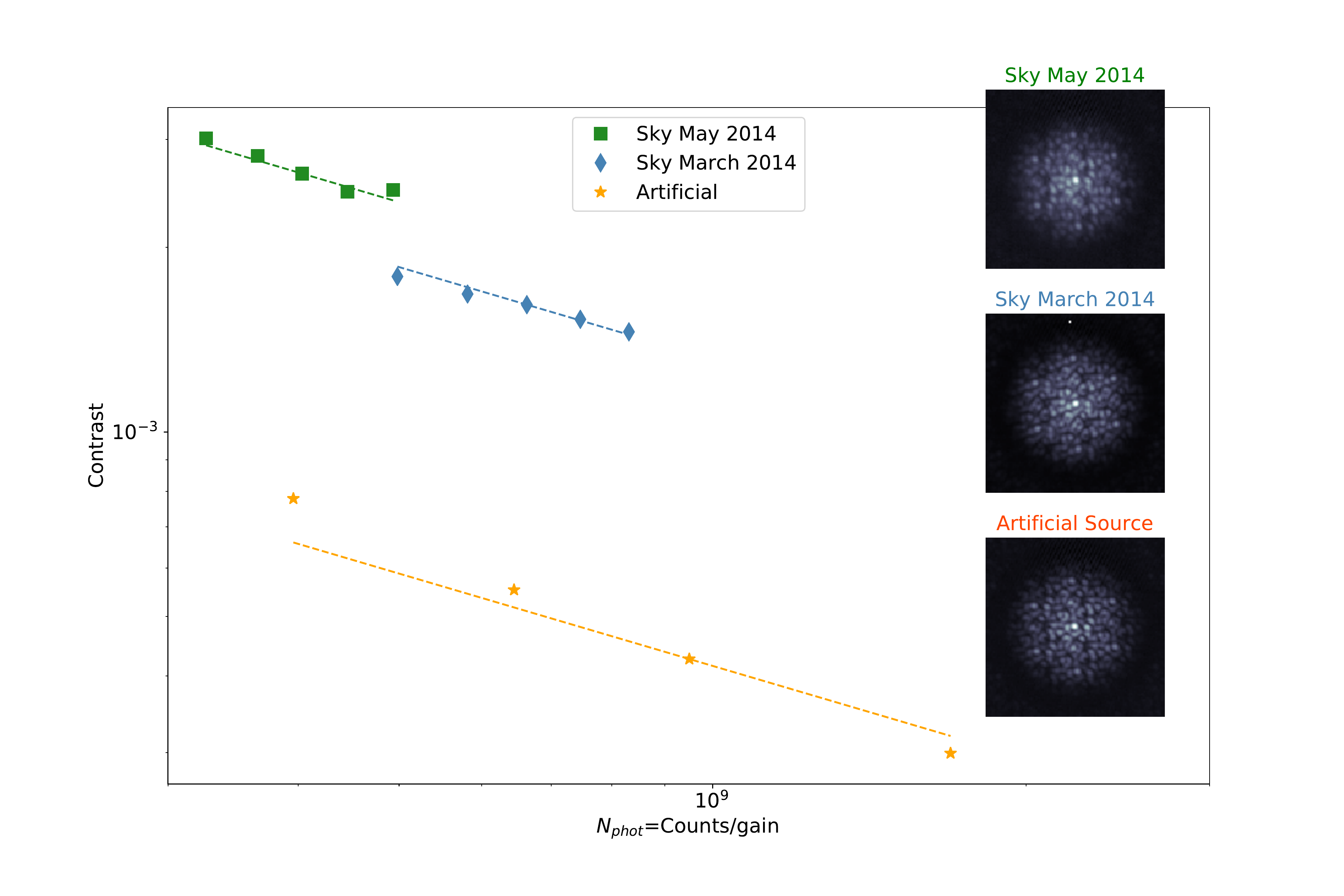}
    \caption{A summary of the median binary contrast sensitivity between
$50-300$ mas for 3 datasets, on-sky observations of HD63852 during two
different observing runs, and internal source exposures. The March observations
showed significantly lower residual AO wavefront error, windspeed, and DIMM
seeing. In better conditions we see both smaller phase errors
and a sharper image, shown on the right inset plots. Table \ref{tab:obs}
shows a more complete list of environmental measurements.
Plotted are the contrast sensitivity obtained with 6, 7, 8, 9
and 10 frames for the on-sky datasets, and 6, 10, 15, and 27 frames for the
internal source dataset. The contrast values are plotted against
total cumulative photon count in the corresponding frames based
on a gain factor of 3.04.  Dashed lines represent a $1/\sqrt{t}$ trend to
compare with the measured contrasts.
}
    \label{fig:mastercontrast}
\end{figure*}

In this section we discuss contrast sensitivity with respect to photon noise
and varying conditions, and provide expected performance for future
observations.  In the best case, images taken with the internal source do not
suffer atmospheric aberration and represent a baseline for performance.
We expect these data to be primarily limited by photon and
detector noise. On-sky observations will suffer from additional aberrations and
smearing out of the image depending on weather conditions.
Observations of an unresolved single star at two different times with different
seeing and wind conditions provide an example of how performance can vary with
conditions. We observed single star HD 63852 on two different nights in H band.
As before, to obtain a proxy for calibrated contrast, we split each sequence of
exposures in half and calibrate the first half against the second half. This
likely overestimates the contrast sensitivity because it
assumes no phase error differences between the target and calibrator. However,
this exercise demonstrates trends in contrast performance with various
environmental conditions and represents an ideal case.  In a
full science sequence one or more different unresolved sources will be used to
calibrate the science target. Calibrators lie in different parts of the sky and
the observations are separated in time between slew and acquisition. This leads
to imperfect correction of closure phase errors. 

In practice NRM contrast will be limited by a range of factors other than
photon noise. Uncharacterized detector noise, vibrations, and
imperfect AO correction that lead to smearing of fringes during an exposure
integration can contribute to reduced contrast.  To characterize the
performance, for each set of observations, we measure closure phases and
scatter with increasing photon count by analyzing partial datasets at a time,
adding in consecutive exposures to increase total counts.  In Fig.
\ref{fig:mastercontrast} we display the measured binary detection sensitivity
against photons collected (detector counts divided by the
recorded gain factor). We compare the measured contrast with a  $1/\sqrt{t}$ 
trend and see some deviations that indicate other systematic errors in closure
phase.

All dataset contrasts improve with increased exposure time
but on-sky observations are not photon noise limited. The dominant error
source in this case is likely time-varying aberrations and vibrations that
reduce fringe visibility (smear out the PSF), resulting from a range of weather
conditions that control the atmospheric turbulence times scale. Systematic
errors are known to limit performance (phase errors) in aperture masking data
\citep{lacour2011}.

The first flux ratio minimum (H-band) is at $40 \mathrm{mas}$. To compare, we
report the average contrast  measured between 100 and 300 mas for each dataset.
For images taken with the internal source, contrast improves with increased
exposure time following the photon noise limit $\sim\sqrt{N_{phot}}$.  In a
range of sky conditions, we see that other effects limit contrast. In very good
conditions we find contrast sensitivity at SNR=5 close to
$\Delta \mathrm{mag}$=7.5 at separations greater than $40 \mathrm{mas}$. We
found that in conditions with higher wind and low level turbulence we measure
an order of magnitude reduced contrast sensitivity for the same bright source.
These conditions generally correspond to Gemini Observatory \textit{IQANY}
conditions with high wind.

\begin{figure}
\centering
    \includegraphics[width=3.4in]{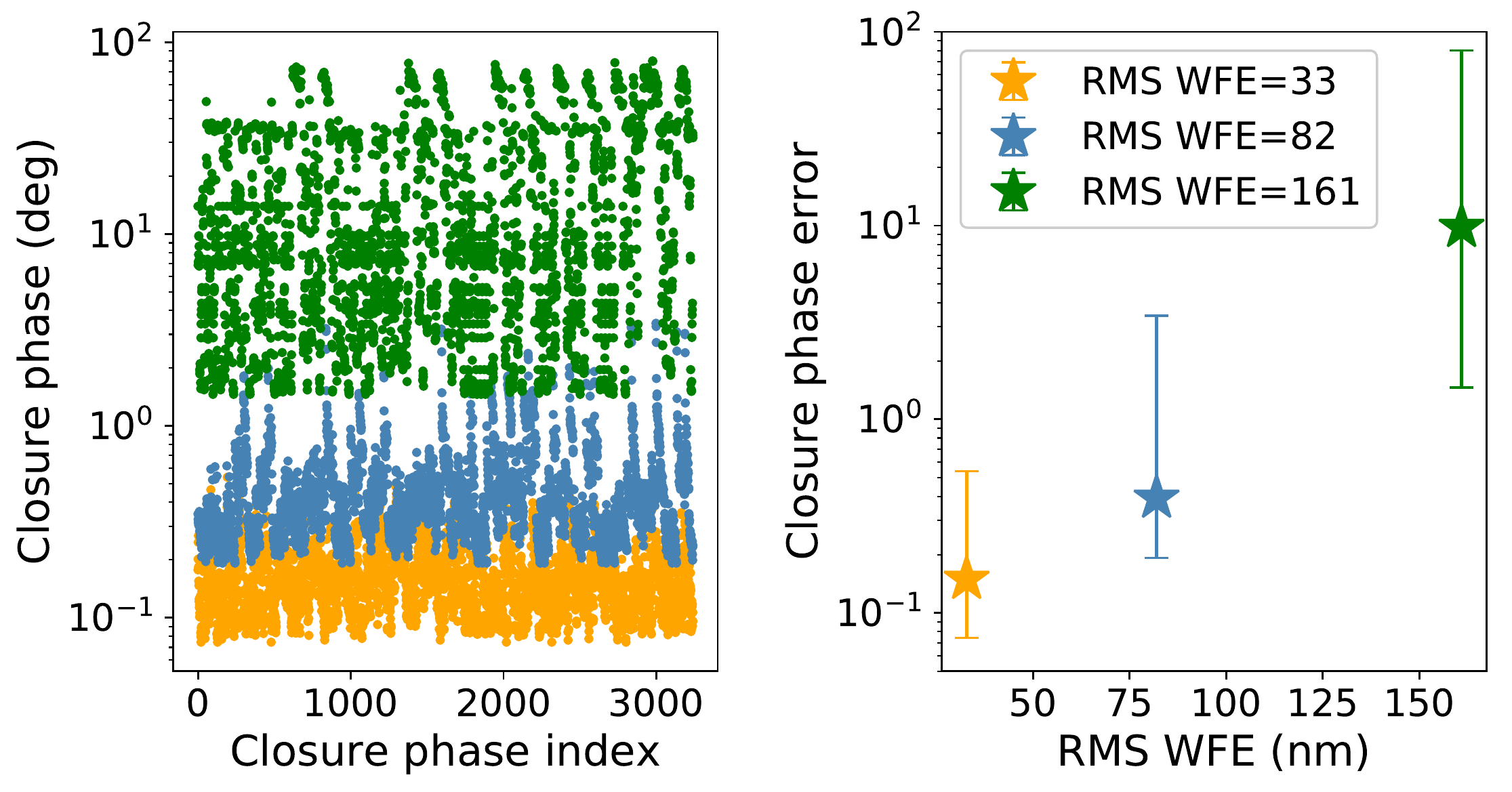}
    \caption{Phase errors compared to residual AO wavefront error (nm) for the
three sets of observations compared in Figure \ref{fig:mastercontrast}. The
left panel shows the error in each  closure phase for every wavelength channel,
while the right panel shows the RMS wavefront error measured from the wavefront
sensor compared to the average closure phase error. Errorbars show the full
range of closure phase errors measured for the dataset. }
    \label{fig:aowfe}
\end{figure}

With few datapoints it is challenging to conclusively identify the dominant
effect reducing fringe contrast, but there are a few obvious correlations. We
note that residual AO wavefront error is a good predictor of point source
contrast. In Figure \ref{fig:aowfe} we show closure phase error as a function
of the wavefront error value reported in the data headers. The cyclical nature
of the phase errors follows a rough scaling with wavelength \citep[also shown
in ][]{greenbaum2014spie}. We also see a correlation with wind speed, however,
the wind speed recorded in the header refers to surface-layer wind and does not
provide any information on wind speed at other levels of the atmosphere.
\cite{madurowicz2018} show that the wind butterfly aberration seen by GPI's
coronagraph~\citep{poyneer2016} most strongly correlates with wind at high
altitudes. It is possible the higher altitude wind was also present during
these observations, or that the ground-layer wind correlates with short
characteristic timescales of atmospheric seeing, also shown to have a strong
effect on GPI performance \citep{bailey2016}.

On-sky observations of fainter targets not only reduces the number of photons
collected, but contains more PSF jitter due to uncorrected wavefront and small
changes in the PSF and/or uncorrected tip/tilt. This has the additional effect
of blurring the image and reducing fringe contrast. This effect is strongest in
poor conditions and especially high winds.

\subsection{Resolving close binary HR2690}
\label{sec:hr2690}

For basic validation of using the NRM to resolve point sources and obtain
precise astrometry we observed the known binary HR 2690 during early
commissioning of GPI. The primary HR 2690A is classified as a B3
star~\citep{buscombe1969}. The contrast ratio of the companion has been
typically measured $\Delta\mathrm{mag}\sim2$ at 0.543$\mu$m \cite{mason1997}.
We observed the binary in the sequence Target-Calibrator-Calibrator. We measure
a contrast sensitivity of $\sim 5 \times 10^{-3}$ by calibrating our two single
stars with each other.

We easily recover the binary in H band and measure a primary to secondary flux
ratio of $5.7 \pm 0.05$ ($\Delta \mathrm{mag}\sim1.89$) a separation of
$89.15 \pm 0.12$ mas, and position angle of $192.29 \pm
0.14^\circ$, after adding GPI plate scale and PA errors in quadrature. 
We find a slight spread in results depending on using one vs. both calibrators,
within the errors. 

\begin{figure}
    \centering
    \includegraphics[width=3.3in]{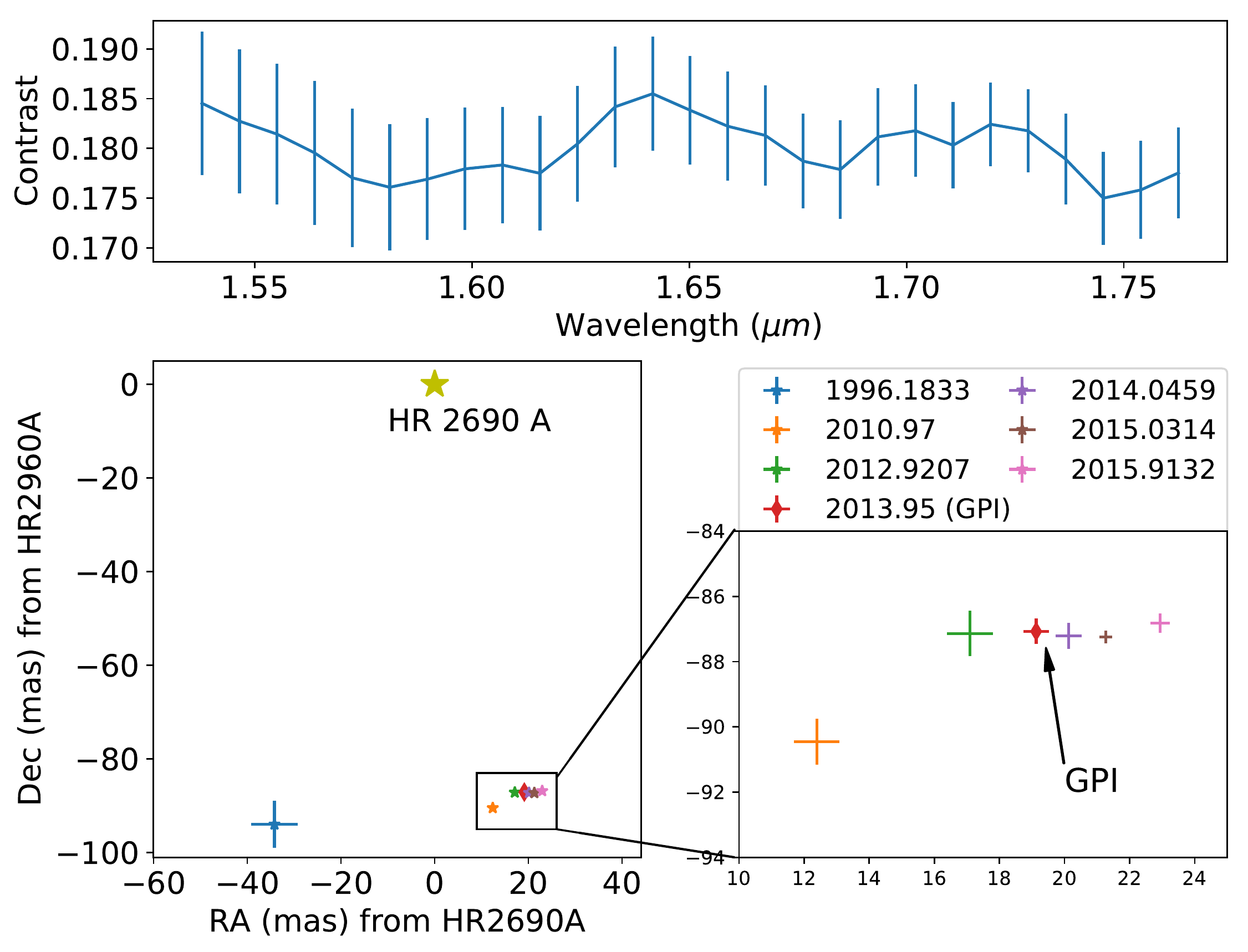}
    \caption{HR 2690 B Recovery. \textbf{Top}: Spectrum (contrast) of the HR
2690 companion measured as the ratio of the secondary to the primary. 
    \textbf{Bottom}: Astrometry of HR 2690 including our GPI epoch. The yellow
star marks the position of HR 2690 A.}
    \label{fig:hr2690astrometry}
\end{figure}

HR 2690 B was first resolved by \cite{mason1997} with speckle imaging. These
observations were followed up several times over the next 19
years~\citep{hartkopf2012, tokovinin2014, tokovinin2015, tokovinin2016}, all
using speckle interferometry. We show the current astrometric positions
relative to the primary including the GPI epoch in Figure
\ref{fig:hr2690astrometry}. The GPI astrometry appears to be consistent with previous measurements. Small discrepancies in astrometry could point to a
mismatch in absolute calibration.

Following the procedure outlined in \S\ref{sec:spect_procedure}, we fit
astrometry and contrast in each wavelength channel. We find a fairly flat
contrast spectrum over H band at $\Delta mag\sim1.89$, which matches the
reported $\Delta mag$ (Stromgren y filter at 0.543$\mu$m) from most of the
previous studies~\citep{mason1997, tokovinin2014, tokovinin2015,
tokovinin2016}. \cite{hartkopf2012} report $\Delta mag_y$=3.2, which is
inconsistent with all other measurements. The similar flux ratio seen at both
visible and near-IR wavelength indicate that the companion is also a hot star,
probably late-B type given these contrast ratios. 

\begin{figure}
    \centering
    \includegraphics[width=3.3in]{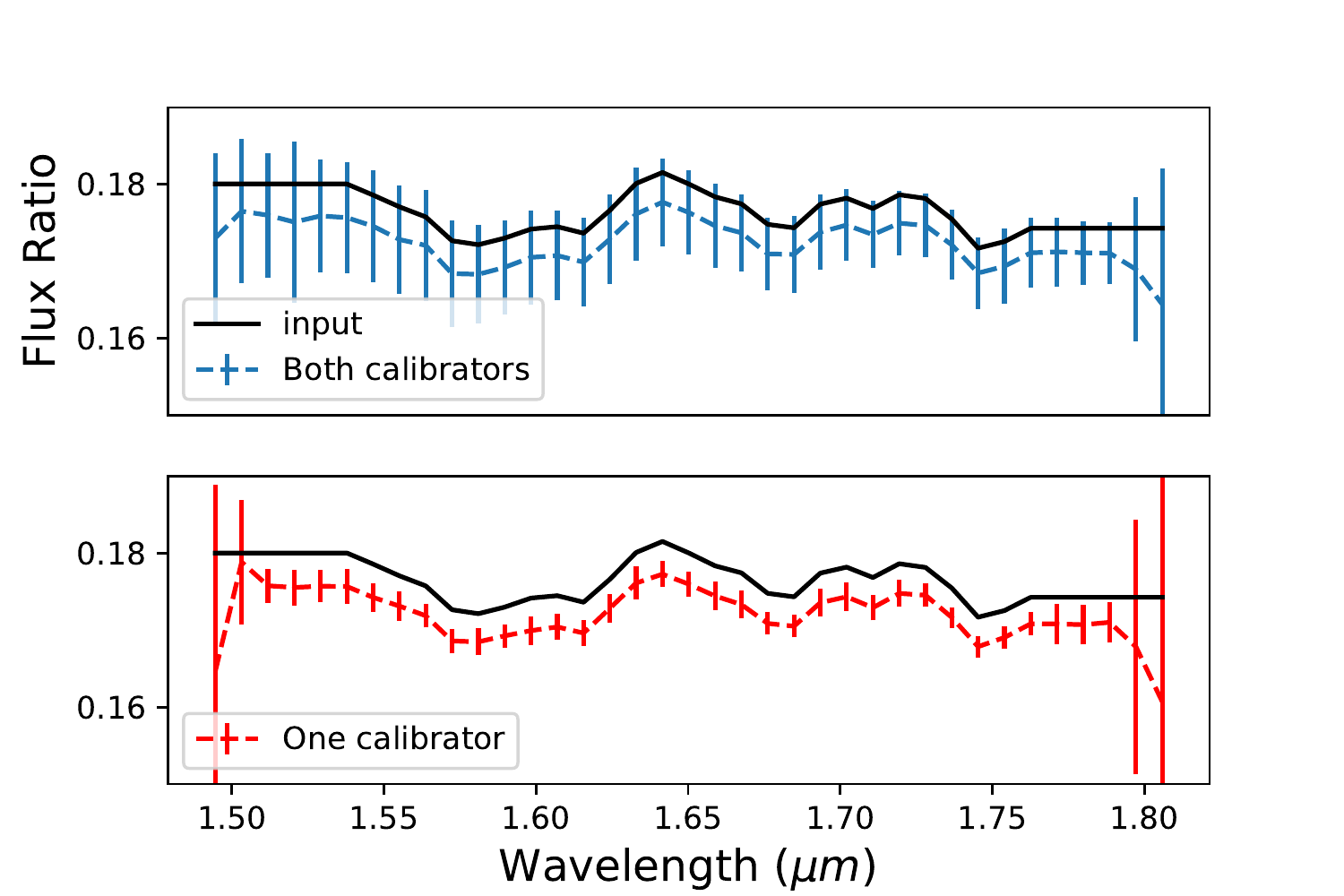}
    \caption{Injection recovery results when \textbf{Top}:
simulated signals are injected into one calibration source and the data are
calibrated using the second PSF calibrator.  and
    \textbf{Bottom}: simulated signals are injected into one calibration
source, and the same original source is used as the calibrator in the analysis.
This simulates the case when only one useable calibration source is available
for injection recovery analysis. In the first case the error in the recovered
spectrum is consistent with $1-\sigma$ errors computed from the MCMC analysis
for the spectrum, but there is a slight bias in the recovered contrast to
smaller flux ratio.  In the second case the errors are underestimated. the bias
is consistent in both cases.}
    \label{fig:hr2690injection}
\end{figure}

As an independent check on our errorbars, we recover simulated
signals using the two calibration sources. We inject and recover a signal into
one of the calibration sources, HR 2839, and use the other, HR 2716, as a sole
PSF calibrator.  We simulate 10 datasets at different position angles near the
separation recovered with the contrast ratio spectrum extracted from the HR
2690 binary. We follow the complete extraction procedure for each simulated
dataset and compute the average and standard deviation. The errors computed by
this approach, shown in Figure \ref{fig:hr2690injection} (top), are consistent
with $1\sigma$ errorbars computed in the original extraction.  In this case
there is also a slight bias in the recovered spectra to lower flux ratio, a
factor $\sim2-3\%$.  For the position we compute a slightly higher error of
$0.4~\mathrm{mas}$ and $0.4^\circ$ for separation and PA, respectively.
The PA shows no strong bias, but the average recovered separation is
approximately $0.4$~mas deviant from the input separation.

In some cases, only one PSF calibrator may be available so it
is not possible to simulate a dataset that accounts for phase errors between
sources. To highlight the difference, we repeat the injection recovery
simulation by calibrating the simulated binary from HR 2839 data with the
original HR 2839 data. As expected, the recovery errors are underestimated.
The contrast ratio spectrum recovered in this simulation is
shown in Figure \ref{fig:hr2690injection}. Interestingly, both the 2-calibrator
simulation and this 1-calibrator simulation show the same ``bias" in the
recovered spectrum (shifted by $2-3\%$). In the case that only one calibration
source is available, injection recovery can be used to measure a systematic
offset in the parameters. 

Using errors computed through injection recovery and mulitplying
by the computed bias term we show the
final spectrum and astrometry of our GPI epoch of observation in Figure
\ref{fig:hr2690astrometry}.  The flat spectrum over this short range is
consistent with a late B-type companion. GPI NRM relative astrometry
measurements are consistent with other high resolution observations and can
reach precision of $\sim0.5~\mathrm{mas}$ in separation and
$\sim0.5^\circ$ in PA.

\subsection{Resolving M dwarf companion inside the transitional disk of HD 142527}
\label{sec:142527}

\begin{figure*}[t]
    \centering
    \includegraphics[width=5in]{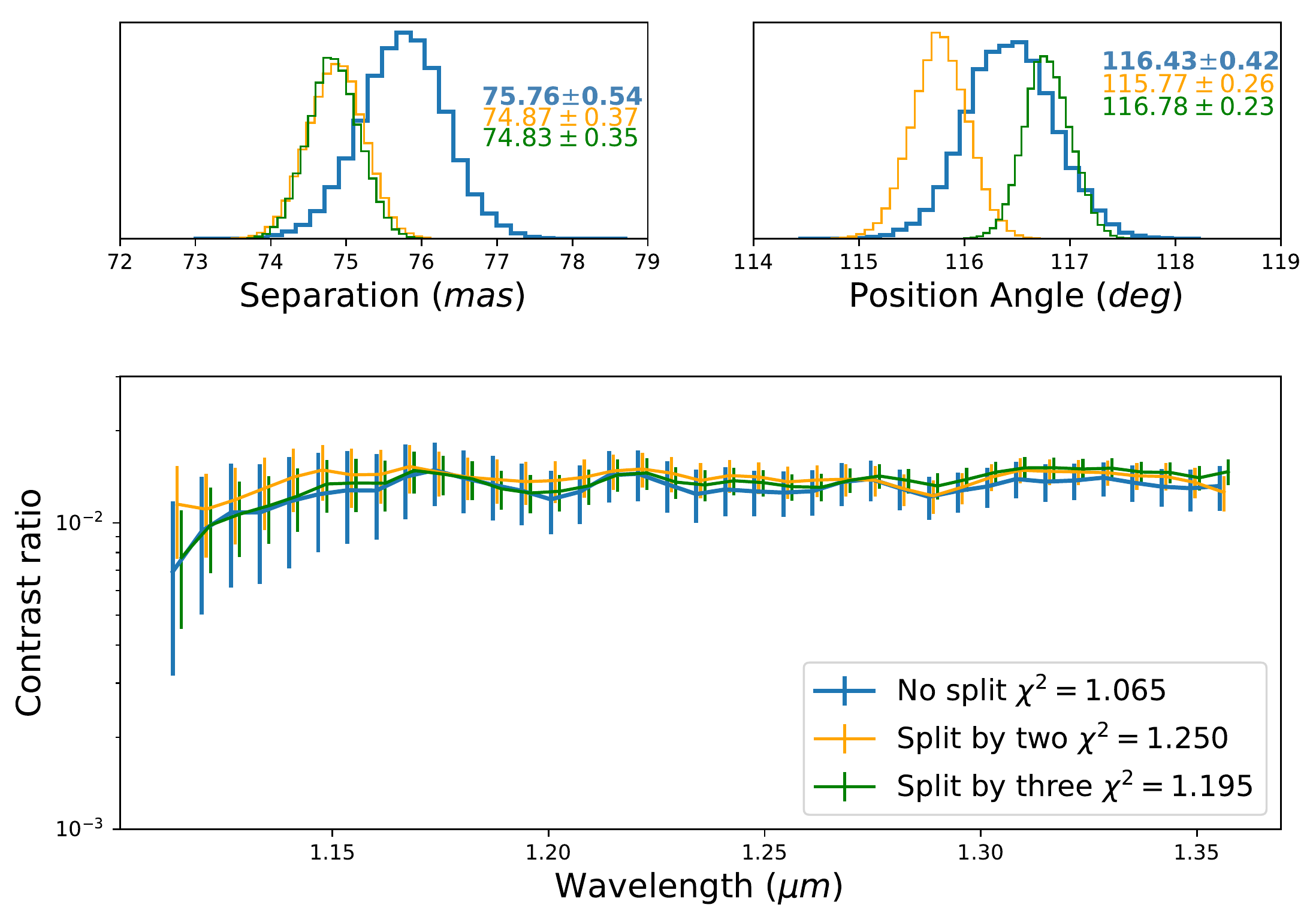}
    \caption{Simultaneous recovery of relative position and
contrast ratio spectrum for three cases that use all 9 frames
of data. The blue curves denote results obtained taking the
average obervables over all 9 frames, considering the average baseline (average
parallactic angle) over the observing sequence. The orange and green curves
represent results when splitting the data by two and three parts, respectively,
and combining/stacking those datasets, thus accounting for sky rotation over
the observing sequence. The results were computed by adding $0.5\circ$
additional phase error in quadrature to the closure phase observables. 
\textbf{Top:} Posteriors for position parameters in each case.  Errorbars
reported are $1\sigma$ (not including GPI astrometry
errors). All approaches, using the same updated calibration favor smaller
separation and discrepant PA than the initial data reduction. The split cases,
while producing tighter errorbars, are not consistent with each other, and lead
to a poorer fit of the data. \textbf{Bottom}: The resulting
contrast spectrum is consistent for each approach. Individual data points are
slightly offset to display relative errorbars. Comparison of reduced $\chi^2$
shows that using the average of all the data provides the best fit in this
case.} 
    \label{fig:hd142527}
\end{figure*}

To demonstrate GPI NRM performance for detection and characterizion of faint
companions at small angular separations, we observed the transitional
disk-hosting, close binary system HD 142527. These data were first presented in
\cite{lacour2016}. We present a new analysis here with more detail and compare
the new spectrum in J-band to photometry and spectroscopy from other
instruments. 
Since the second calibration source was determined to be a
close binary~\citep{lebouquin2014} we only have one calibration source
available for this analysis and the complete injection recovery approach to
estimating errorbars is not possible. We perform the injection recovery to
reveal any extraction biases, relying on $1\sigma$ errorbars computed by the
MCMC reduction algorithm, which we have shown to be consistent with injection
recovery errors in the previous example. Extraction biases are $\sim5\%$, which
we apply to the resulting spectrum for \S\ref{sec:hd142spectrum}.

\subsubsection{Recovering parameters \label{sec:hd142params}}

We first determine the position and average contrast ratio. We follow a similar
procedure reported in \cite{lacour2016}, using all frames except two where the
AO system lost lock on the star and the images are noticeably blurred. We use
an average sky rotation and consider the whole dataset at this common
parallactic angle.  This position is in agreement with previously measured
astrometry~\citep{biller2012, close2014, lacour2016}.  We measure an average
contrast ratio of $\sim70$ between the primary and the secondary, which is
consistent with measurements in J and H bands~\citep{lacour2016} with NACO
sparse aperture masking.  Next we compute the full set of parameters, contrast
ratio for each wavelength channel and position, as outlined in
\S\ref{sec:spect_procedure} adding $0.5^\circ$ additional closure phase error
in quadrature.  We obtain a projected separation of $75.76
\pm0.54~$mas and PA of $116.43 \pm0.44^\circ$ ($\chi^2=1.07$).

\begin{figure*}
    \centering
    \includegraphics[width=5in]{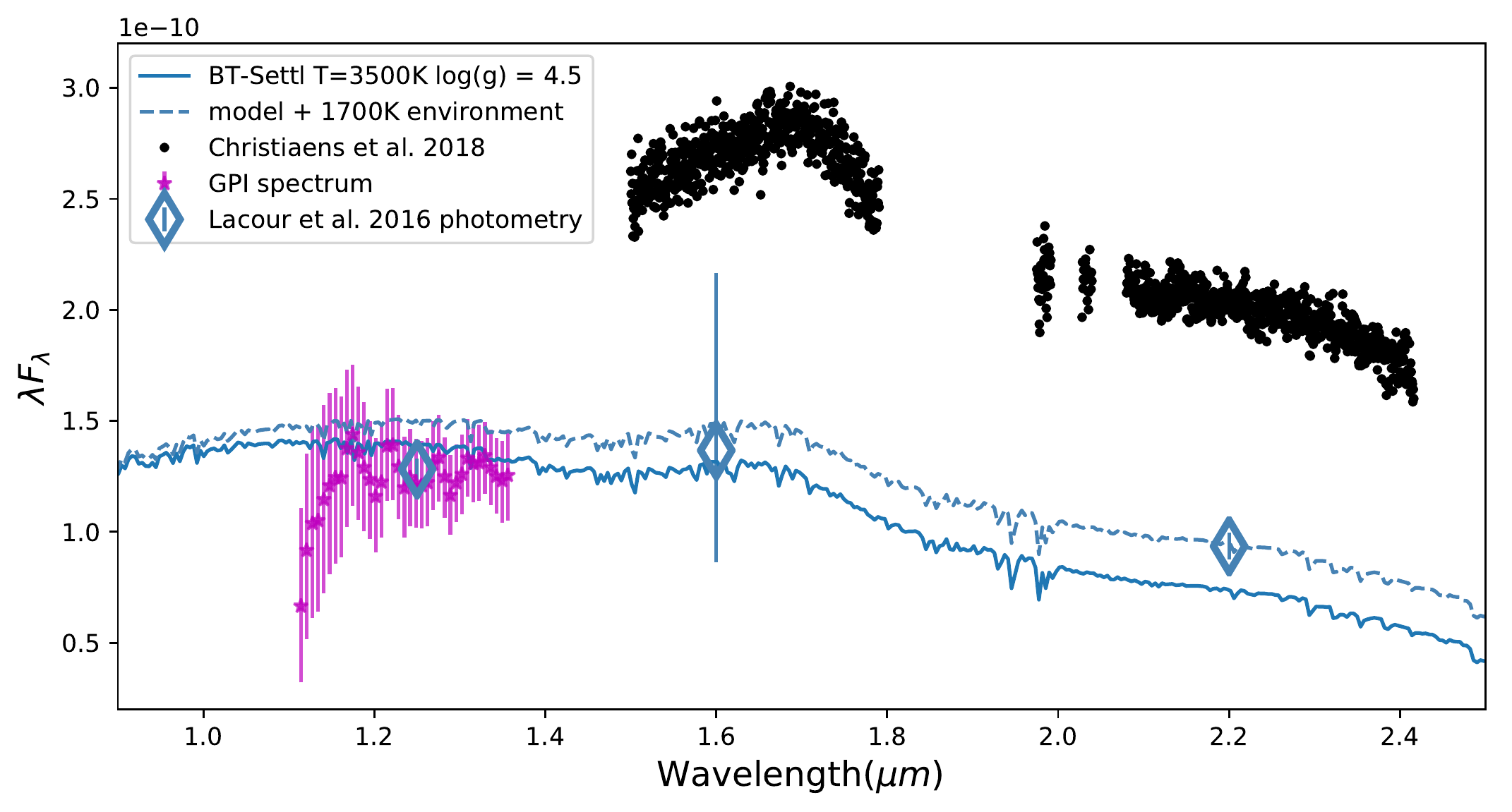}
    \caption{HD 142527 B spectrum converted to flux based on the host star
photometry (blue diamonds) from \cite{lacour2016}.  Blue lines represent model
spectra described in \cite{lacour2016} and \cite{christiaens2018} with
$T=3500\mathrm{K}$, $log(g)=4.5$, alone (solid) and with a 1700K environment
(dashed), assuming a distance of 140pc. Our spectrum of HD 142527 B is
consistent with this previously measured photometry, which are discrepant from
new VLT/SINFONI H+K spectroscopy~\citep{christiaens2018} displayed in black
dots.}
    \label{fig:hd142527_context}
\end{figure*}

Since there is a significant amount of sky rotation (11.4$^\circ$) over the course
of the HD142527 integrations, we explored the effect of
splitting the dataset into two groups of four and five
exposures and three groups of three exposures, combining the rotated baselines
in the analysis. We refer to this as the ``split and combine"
method. In this case the contrast ratio between HD142527 A and
B is slightly higher.  The parameter errorbars are also slightly smaller due to
the larger number of observables.  We recover slightly smaller separations of
$74.83 \pm0.35~$mas and $74.87\pm0.37~$mas and discrepant PAs of $115.8
\pm0.29^\circ$ and $116.8 \pm0.26^\circ$ for the split in two ($\chi^2=1.25$) and
split in three ($\chi^2=1.19$) cases.

Next we use the three-parameter analysis results as a starting
guess to simultaneously fit for position and a contrast ratio in each spectral
channel for each of the three reduced datasets, the average of all frames, and
the split and combined by two and three. The comparison is shown in Figure
\ref{fig:hd142527}. The two split datasets still produce a smaller separation
and discrepant PAs. However, all reduced datasets produce a consistent
spectrum.  A known degeneracy between separation and contrast could be the
cause of a smaller recovered separation, but the discrepancy in PA is likely
due poor data quality, since the results depend on how the data are
combined. The small number of total frames makes this approach challenging.

While there is some variation in the position parameters, there is not a large
difference in the spectrum of each reduction within the errorbars. We adopt the
solution with the lowest error between the data and model (lowest $\chi^2$).
Obtaining reliable astrometry may require more integrations in order to average
out poor quality data and get a cleaner picture of the true astrophysical
structure. 

\subsubsection{The HD 142527 B spectrum  \label{sec:hd142spectrum}}

H$\alpha$ was previously detected at visible wavelengths~\citep{close2014},
however, given our low resolution spectrum, our errors are too large to see the
Pa$\beta$ signal expected accretion luminosity reported in either
\cite{close2014} (1.3\%~$L_\odot$) or \cite{christiaens2018} (2.6\%~$L_\odot$).
The expected line luminosity is approximately an order of magnitude smaller
than our errorbars, according to the relations, \begin{eqnarray}
\log{(L_{acc})} &=& B + A \times L_{line} \\ A_{Pa\beta} &=& 1.36,
B_{Pa\beta}=4.00 \end{eqnarray} as described in \cite{natta2004, rigliaco2012}.

We correct our recovered spectrum with the $\sim5\%$ extraction
bias factors computed from injection recovery in the calibration source
dataset. The recovered spectrum is consistent with the broadband photometry
previously measured for HD 142527 B~\citep{biller2012,lacour2016,close2014}.
Figure \ref{fig:hd142527_context} shows our J band spectrum next to published
photometry (blue diamonds).  We overplot a $T_{eff}=3500\mathrm{K}$ model alone
and one with a 1700K environment (similar to the models described in
\cite{lacour2016} and \cite{christiaens2018}), assuming a distance of 140~pc to
be consistent with \cite{lacour2016}. Our results are consistent with the
aperture masking detections.  We also plot the higher resolution VLT/SINFONI
H+K spectra from \cite{christiaens2018} (black dots), and note the flux
discrepancy. 
The discrepancy with \cite{christiaens2018} is most likely a systematic error
in one or both of the analyses. The presence of bright extended structures
could bias the recovery of the secondary point source position and flux, but a
point source was also detected in direct imaging~\citep{close2014}. Our
results, taken independently, support previous aperture masking measurements,
and we have demonstrated that our analysis procedure yields reliable
measurement of the spectrum in simulations. Alternatively, it is possible that
inaccurate calibration of the SINFONI data in post-processing could yield this
discrepancy.
The stellar spectrum models described in both studies assumed difference
distances for HD 142527 B, $140\pm20~$pc~\citep{lacour2016} and
$156\pm6~$pc~\citep{christiaens2018} resulting from the parallax measured with
Gaia \citep{gaiacollab2016}.  We note the coincidence that the flux discrepancy
is close to the scaling factor between these distances ($156^2/140^2$). If the
deeper contrast measured from this and other aperture masking observations are
correct, this may imply a lower effective temperature, or different
circumbinary environment. 
The small separation of HD 142527 B makes non-coronagraphic, full pupil images
challenging to reduce.

\section{Polarimetric mode \& visibility precision}
\label{sec:pol}

Reliable visibility amplitudes are challenging to measure from the ground, even
behind an extreme-AO system. Small temporal changes in phase smear fringes over
individual integrations and vibrations artificially reduce amplitudes.
Differential polarimetry enables self-calibrated amplitudes under the
assumption that orthogonal polarization channels and rotated half-waveplate
angles are expected to suffer the same systematic errors. These systematics can
therefore be calibrated out to reveal different polarized structure. In this
section we follow the polarimetry+NRM procedure outlined in Sec
\ref{sec:observables} and report performance of the NRM in polarimetric mode.

In commissioning the polarimetric mode we focused on single, unresolved
calibration stars. The differential visibility signal is expected to be
unpolarized and should show constant $\mathcal{V}_{diff}=1$ and $\phi_{CP} = 0$
at all orientations. The deviation from the expected signal and scatter provide
an estimate of both instrumental systematics and stability of the measurements.
During commissioning observations in May 2015, when we experienced large
vibrations, differential visibilities had very large errorbars and residual
systematic scatter around $\mathcal{V}_{diff}=1$. Vibrations were exacerbated
by high winds during May 2015 NRM commissioning.

\begin{figure}
    \centering
    \includegraphics[width=3.3in]{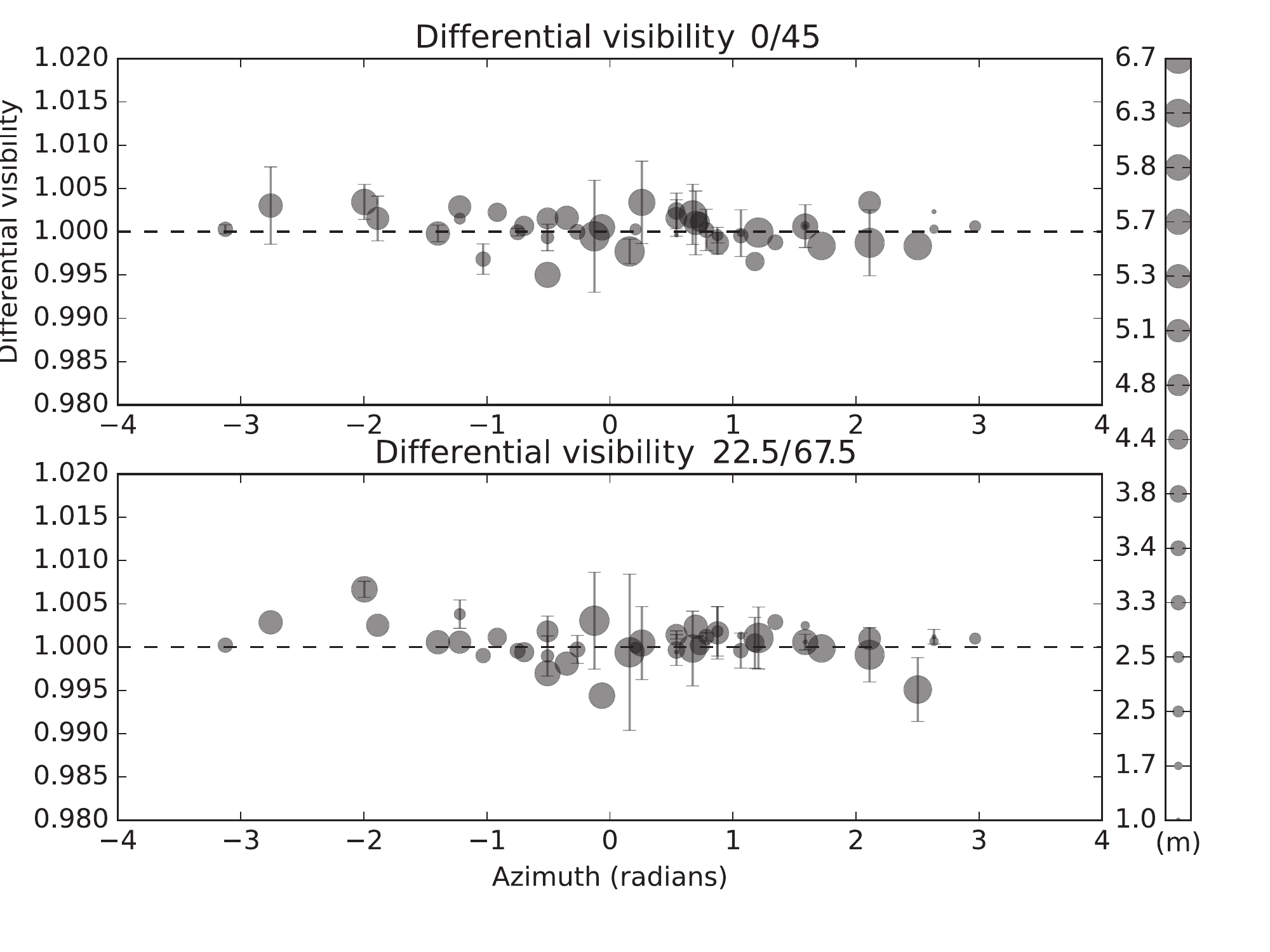}
    \includegraphics[width=3.3in]{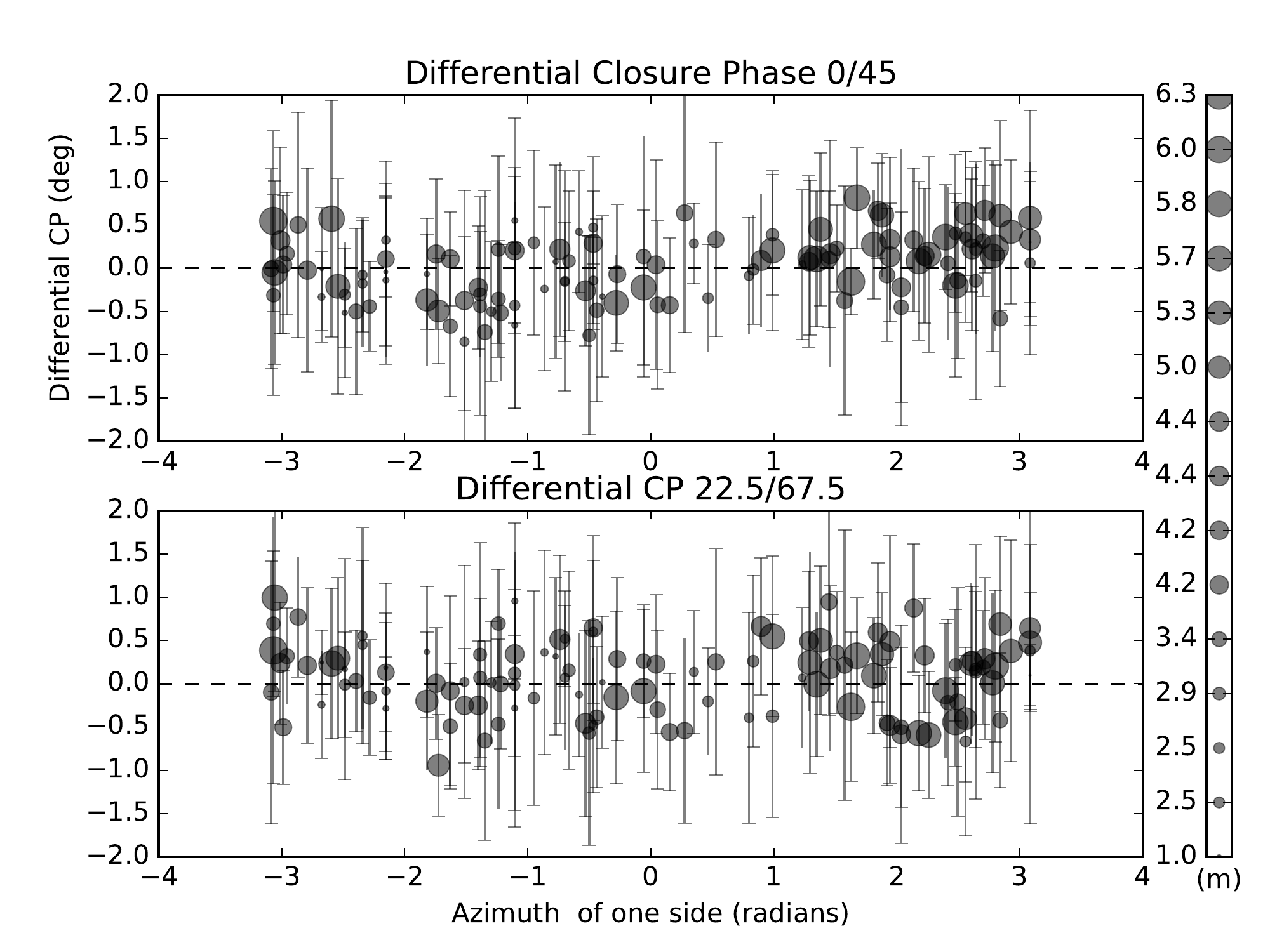}
    \caption{Differential visibilities (top) and differential closure phases
(bottom) for a representative polarimetric dataset of a single, unpolarized
star. Marker size scales with baseline length. We plot the differential
visibility with baseline azimuthal orientation.}
    \label{fig:polperformance}
\end{figure}

In May 2016 commissioning, after a major source of vibration was fixed, we
found that, in the best cases, differential visibilities calibrated to within
$1\%$ of $\mathcal{V}=1$, $\sigma\sim0.4\%$ in the best case
shown here. Closure phases calibrated within $\sim1-2^\circ$ for bright
sources, $\sigma\sim0.4^\circ$ in the best case. For example, Figure
\ref{fig:polperformance} shows the measured differential visibilities for
single source HIP 74604 from data taken in GPI's K1 band.  This
represents the best performance we achieved during commissioning, which is
similar to the $\sigma\sim0.4\%$ performance achieved with VAMPIRES polarimetry
mode at visible wavelengths \citep{norris2015}.

We explored the expected differential polarimetry signal of a protoplanetary
disk by simulating the instrument response for a synthetic disk produced with
MCFOST \citep{pinte2006,pinte2009} and reducing this simulated data through our
pipeline. Within a limited set of tests attempting to simulate a relatively
large signal, we were not able to simulate a detectable disk at the level of
noise we measure from our best on-sky data, without artificially dialing down
the flux from the star by a factor of a few. As an example (described in detail
in Appendix \ref{sec:polsim}), we simulated data based modeling the features of
HD 97048~\citep{lagage2006, doucet2007}, a young Herbig Ae star with a strong
IR excess, $L_{IR}\sim0.4L_\odot$ \citep{vankerckhoven2002}. We physically
scale disk image so that the inner edge of the disk is located $<100~$mas from
the central star. In this example, the integrated flux into one GPI pixel
($14.1~$mas) of the brightest part of the disk inner edge is still $\sim7~$mag
fainter than the host star (see Figure \ref{fig:mcfost} in Appendix
\ref{sec:polsim}).

\section{Discussion}
\label{sec:discussion}

GPI's non-redundant mask mode in general shows comparable performance compared
to prior aperture masking~\citep[e.g.][]{lacour2011} and earlier IFS aperture
masking~\citep{zimmerman2012} experiments, and very good performance in
good conditions that correspond to low residual WFE measured
by the AO system. As \cite{zimmerman2012} showed in the P1640 instrument, the
IFS spectral axis provides improved overall contrast compared to broadband
aperture masking and also smooths out baselines with lower sensitivity. This
allows GPI NRM to reach contrasts close to $10^{-3}$ on bright targets and
better than $10^{-2}$ on long individual integrations ($\sim$20-60 seconds).
GPI's NRM achieves similar performance at $\sim\lambda/D$ in J
and H bands as NACO SAM L$'$ imaging of similar total integration time, which
achieved contrast limits of $2.5\times10^{-3}$ \citep{lacour2011}. Deeper NACO
L$'$ imaging \citep{gauchet2016} exceeds this sensitivity, especially for
bright sources. GPI's 10-hole mask, while reducing throughput compared to
other masks with fewer holes provides fairly even coverage of spatial
frequencies.

We find that, in addition to helping constrain the average contrast
measurement, we can fit a spectrum reliably to moderate contrast sources, with
improved overall contrast. We have presented new spectra of HR 2690 B in H band
and HD 142527 B in J band that are consistent with previous photometry for both
sources. A flux discrepancy remains between aperture masking observations of HD
142527 B and VLT/SINFONI spectra in H and K bands \citep{christiaens2018}.
Future observations may help resolve this discrepancy.  GPI's IFS mode combined
with the NRM is particularly powerful for obtaining precise ($\sim
\mathrm{few~mas}$) astrometry of companions around bright host stars that are
separated $< 100$ mas, where methods like Angular Differential
Imaging~\citep{marois2006} suffer.

The ability to resolve the relative astrometry and
spectro-photometry of close binaries is a valuable tool for studying stellar
multiplicity and calibrating evolutionary models as a function of mass and age.
In addition, determining the mass and SED of both components of binary members
of a moving group can constrain the age of the group as a whole, especially if
a pre main sequence star is moving along the Henyey track (e.g. Nielsen et al.
2017).  The best targets for this technique have short orbital periods to allow
for quick characterization and large radial velocity signals, which in nearby
moving groups means projected separations of $\sim$40 mas.  Typical contrasts
can reach masses $\sim50M_{jup}$ for a very young bright target as we have
shown in Figure 2 considering the AMES-Cond models (Baraffe et al. 2003) for a
1Myr 6.5 mag primary at 140pc.

In a single observing sequence, we obtained target integrations with minimal
sky rotation, where possible, to provide multiple independent measurements
along the same sky-projected baselines. In cases with a larger amount of sky
rotation, we explored splitting up datasets to account for
the rotation and take advantage of the
increased Fourier coverage. While this can reduce the error on the fit, for 
the very small number of frames we obtained this produced discrepant results
depending on how the data were split, sensitive to variations between frames.
Ultimately averaging observables over all frames produced a better fit model 
for the HD 142527 dataset. All approaches yielded consistent spectra, but 
saw some variation in the relative position. With observations covering even
greater sky rotation a split and combine approach will likely be neccesary
and should be robust if the uncertainties on the observables can be estimated
(e.g., by collecting a sufficient number of frames for each sky position). 

Polarization mode observations rely on measuring stable amplitudes, which
become degraded by vibrations and poor wavefront corrections. We saw
improvement in precision after major sources of vibration were corrected. A
faulty M2 mirror actuator was fixed and active dampers were installed. In the
best case of the most recent observations, we measured precision of
$\sigma\sim0.4\%$ in differential visibilities and
$\sim0.4^\circ$ in differential closure phases in the best case. However, our
limited observations make it difficult to characterize the typical polarimetric
mode performance with the NRM on GPI.  Initial attempts to simulate NRM images
of a model protoplanetary disk did not yield a detectable signal; a disk will
need to be relatively bright to be detected. Compared with the
VAMPIRES instrument \citep{norris2015} we reach similar performance in our best
dataset taken in K1 band. Typical VAMPIRES performance is likely better and
their three-tier calibration (compared to GPI's two-tier described in
\S\ref{sec:observables}) makes that system more robust to systematic errors.

At this level of precision, young circumstellar disks may be a significant
challenge to detect with differential polarimetry on GPI, compared to sources
previously detected by this method with larger polarization
signals~\citep[e.g.,][]{norris2012}.  In the case of a resolved signal with
NRM+polarimetry, modeling is an essential component for recovering and
interpreting the disk structure. Studying suspected polarized extended
structures with NRM should be limited to the best conditions (low residual
wavefront error, low wind). Future upgrades or instruments
that can mitigate vibrations and tip/tilt errors for non-coronagraphic modes
could make better use of polarimetry with NRM for studying circumstellar disks.

\section{Summary and Conclusions}
\label{sec:summary}

We have outlined the overall performance of the GPI NRM in IFS and polarimetric
modes with a few example datasets. We have also described an open source
software to reduce NRM fringes from GPI and other instruments and demonstrated
results on various datasets.  Future observations with the NRM on IFS
instruments like GPI can use this study as a guideline for observing in these
modes. 

We also provide the following major takeaways for planning observations with
GPI's NRM: 
\begin{itemize} 
\item AO residual wavefront error correlates with NRM contrast performance
(Fig. \ref{fig:mastercontrast}). The AOWFE header keyword is a good metric of
conditions for NRM performance, given the ``long" integration times.  
\item GPI NRM is suitable for moderate contrasts between $10^2 - 10^3$ to
separations of $\sim30\mathrm{mas}$, with degraded performance closer in.  
\item Ten holes provides good uv coverage minimizing gaps of sampling
sensitivity, but at the cost of lower throughput.  
\item Polarization observations should be taken in conditions that minimize AO
residual wavefront error and when vibrations can be minimized. 
Polarization observations should target objects with differential
polarimetry signals $\gtrsim 1\%$.
\end{itemize}

This study can provide a comparison with other instruments using single-pupil
interferometric methods (ie., NRM, kernel phase~\citep{martinache2010}).
Further improvements to the analyses presented in this work could be made by
analyzing statistically independent kernel phases \citep[i.e.,][]{ireland2013}
or with more sophisticated modeling and treatment of errors. The richness of
the IFS datasets allows for varied approaches to treating and analyzing the
data. 

Ground-based NRM on instruments like GPI complement the capabilities of
upcoming NIRISS aperture masking on JWST. The obvious advantage of NRM on
ground-based facilities like GPI is in the larger telescope size that enables
higher resolution sensitivity down to $10$s of milli-arcseconds. On the other
hand, interferometric observations on a stable space telescope like JWST will
carve out a different discovery space. The data will likely be photon-noise
limited for bright sources, allowing at least an order of magnitude improved
contrast compared to the ground. Space-based interferometric
observations will also be able to complement ground-bases AO observations by
observing sources too faint for visible wavefront sensors. Together with other
high contrast and high resolution instruments, IFS aperture masking
observations help to expand the detection landscape for direct imaging.

\acknowledgments
The authors thank Valentin Christiaens for sharing their VLT/SINFONI data and
Neil Zimmerman for useful discussions. We thank the anonymous 
reviewer for helpful comments that improved the clarity of this paper. 
This research has made use of the SVO
Filter Profile Service (http://svo2.cab.inta-csic.es/theory/fps/) supported
from the Spanish MINECO through grant AyA2014-55216. 
 
This work is based on observations obtained at the Gemini Observatory, which is
operated by the Association of Universities for Research in Astronomy, Inc.,
under a cooperative agreement with the NSF on behalf of the Gemini partnership:
the National Science Foundation (United States), the National Research Council
(Canada), CONICYT (Chile), Ministerio de Ciencia, Tecnolog\'{i}a e
Innovaci\'{o}n Productiva (Argentina), and Minist\'{e}rio da Ci\^{e}ncia,
Tecnologia e Inova\c{c}\~{a}o (Brazil). 
Work from A.Z.G was supported in part by the National Science Foundation
Graduate Research Fellowship Program under Grant No. DGE1232825. A.Z.G and A.S.
acknowledge support from NASA grant APRA08-0117 and the STScI Director’s
Discretionary Research Fund.
The research was supported by NSF grant AST-1411868 and NASA grant NNX14AJ80G
(J.-B.R.).
P.K., J.R.G., R.J.D., and J.W. thank support from NSF AST-1518332, NASA
NNX15AC89G and NNX15AD95G/NEXSS.  This work benefited from NASA’s Nexus for
Exoplanet System Science (NExSS) research coordination network sponsored by
NASA's Science Mission Directorate.  KMM's work is supported by the NASA
Exoplanets Research Program (XRP) by cooperative agreement NNX16AD44G.
Portions of this work were performed under the auspices of the U.S. Department
of Energy by Lawrence Livermore National Laboratory under Contract
DE-AC52-07NA27344.  
\software{Astropy \citep{astropy2013}, Numpy \citep{Numpy2011}, Scipy
\citep{scipy2001}, oifits \footnote{https://github.com/pboley/oifits},
pysynphot \citep{pysynphot2013}, emcee \citep{dfm2013soft, dfm2013}}

\facility{Gemini South}.

\appendix
\section{Contrast spectra of HR 2690 B and HD 142527 B}

We provide the contrast spectrum of HR 2690 B in Table
\ref{tab:hr2690spect} and HD 142527 B in Table \ref{tab:hd142527spect}.
Absolute flux calibrations depend on the host star photometry and choice of
model spectrum. 

\begin{deluxetable}{l l l l }

\tablecaption{Flux ratios recovered for HD 2690 B. \label{tab:hr2690spect}}
\tablehead{\colhead{Wavelength ($\mu\mathrm{m}$)} & \colhead{Flux Ratio} & 
\colhead{$\mathrm{Error_+}$} & \colhead{$\mathrm{Error_-}$}}
\startdata
\hline
1.537827643	&0.180051769	&0.006117792	&0.005922433 \\
1.546472049	&0.178603929	&0.005840979	&0.006013181 \\
1.555116455	&0.177078709	&0.005528977	&0.005831117 \\
1.563760975	&0.175766975	&0.005111043	&0.005116699 \\
1.572405381	&0.172665394	&0.005382201	&0.005327732 \\
1.581049787	&0.172154242	&0.005318425	&0.005348661 \\
1.589694307	&0.173011673	&0.005209509	&0.005290487 \\
1.598338713	&0.174184211	&0.005493067	&0.005438636 \\
1.606983119	&0.174497753	&0.005225567	&0.005139819 \\
1.615627639	&0.173650569	&0.004966511	&0.004970789 \\
1.624272045	&0.176633443	&0.004756046	&0.004828567 \\
1.632916451	&0.180121456	&0.004977992	&0.004896766 \\
1.641560971	&0.181550508	&0.00510479	    &0.005205588 \\
1.650205377	&0.180051944	&0.005127553	&0.00515305 \\
1.658849783	&0.178376985	&0.004634255	&0.004762568 \\
1.667494303	&0.177465747	&0.004793606	&0.004839009 \\
1.676138709	&0.174799634	&0.004592341	&0.004651037 \\
1.684783115	&0.174348035	&0.004506974	&0.004584267 \\
1.693427635	&0.177431085	&0.004909268	&0.004794985 \\
1.702072041	&0.178212043	&0.00503311	    &0.004966011 \\
1.710716447	&0.176859468	&0.005078732	&0.004995903 \\
1.719360966	&0.178652942	&0.004858403	&0.004780304 \\
1.728005373	&0.178172915	&0.004585516	&0.004590963 \\
1.736649779	&0.175487526	&0.004325864	&0.004293816 \\
1.745294298	&0.171700648	&0.00460216	    &0.004354205 \\
1.753938704	&0.172556787	&0.004636444	&0.004680254 \\
1.762583111	&0.174288292	&0.004666991	&0.004781194 
\enddata
\end{deluxetable}

\begin{deluxetable}{l l l l }

\tablecaption{Flux ratios recovered for HD 142527 B. \label{tab:hd142527spect}}
\tablehead{\colhead{Wavelength ($\mu\mathrm{m}$)} & \colhead{Flux Ratio} & 
\colhead{$\mathrm{Error_+}$} & \colhead{$\mathrm{Error_-}$}}
\startdata
\hline
1.114073029	&0.007267947	&0.004835427	&0.003762216 \\
1.120800789	&0.010011145	&0.004738789	&0.004379546 \\
1.127528549	&0.011327441	&0.004793418	&0.004649354 \\
1.134256308	&0.011465136	&0.004688196	&0.004493155 \\
1.140984068	&0.012490109	&0.00464006	    &0.004623279 \\
1.147711941	&0.01316459	    &0.004552897	&0.004364929 \\
1.154439701	&0.01355351	    &0.004379782	&0.004214281 \\
1.161167461	&0.013537049	&0.004000996	&0.003915397 \\
1.167895221	&0.014960988	&0.003871995	&0.003850851 \\
1.17462298	&0.015653951	&0.003427684	&0.003493358 \\
1.18135074	&0.014759454	&0.003214467	&0.003288693 \\
1.188078613	&0.014018133	&0.003213441	&0.003124573 \\
1.194806373	&0.013425199	&0.002869815	&0.002900563 \\
1.201534133	&0.012590288	&0.002839904	&0.002744232 \\
1.208261892	&0.013288886	&0.002762051	&0.002728965 \\
1.214989652	&0.015046037	&0.002809408	&0.002681133 \\
1.221717412	&0.015079263	&0.002792998	&0.002698887 \\
1.228445285	&0.014030106	&0.002514835	&0.002558348 \\
1.235173045	&0.012988723	&0.002492478	&0.002426185 \\
1.241900804	&0.013489881	&0.002361039	&0.00236329 \\
1.248628564	&0.013230286	&0.002148547	&0.002178483 \\
1.255356324	&0.013107222	&0.002199957	&0.002087967 \\
1.262084083	&0.013264332	&0.002177131	&0.002152015 \\
1.268811957	&0.014177584	&0.002082765	&0.002253849 \\
1.275539717	&0.014503863	&0.002086962	&0.00215929 \\
1.282267476	&0.013568453	&0.001929619	&0.002003019 \\
1.288995236	&0.012662013	&0.001967162	&0.001922264 \\
1.295722996	&0.013277692	&0.001895677	&0.001937012 \\
1.302450755	&0.013703473	&0.001951891	&0.001962818 \\
1.309178515	&0.014498995	&0.001914941	&0.001903277 \\
1.315906388	&0.014245619	&0.001920155	&0.001907616 \\
1.322634148	&0.014284923	&0.001862644	&0.001861593 \\
1.329361908	&0.01462621	    &0.001683988	&0.001860377 \\
1.336089667	&0.014070104	&0.001864895	&0.001855666 \\
1.342817427	&0.013637269	&0.001885681	&0.001836921 \\
1.349545187	&0.013433555	&0.0019365	    &0.002086427 \\
1.35627306	&0.013692138	&0.002108377	&0.002229165 
\enddata
\end{deluxetable}

\section{Synthetic polarimetry observation example} \label{sec:polsim}
\begin{figure}
    \centering
    \includegraphics[width=3in]{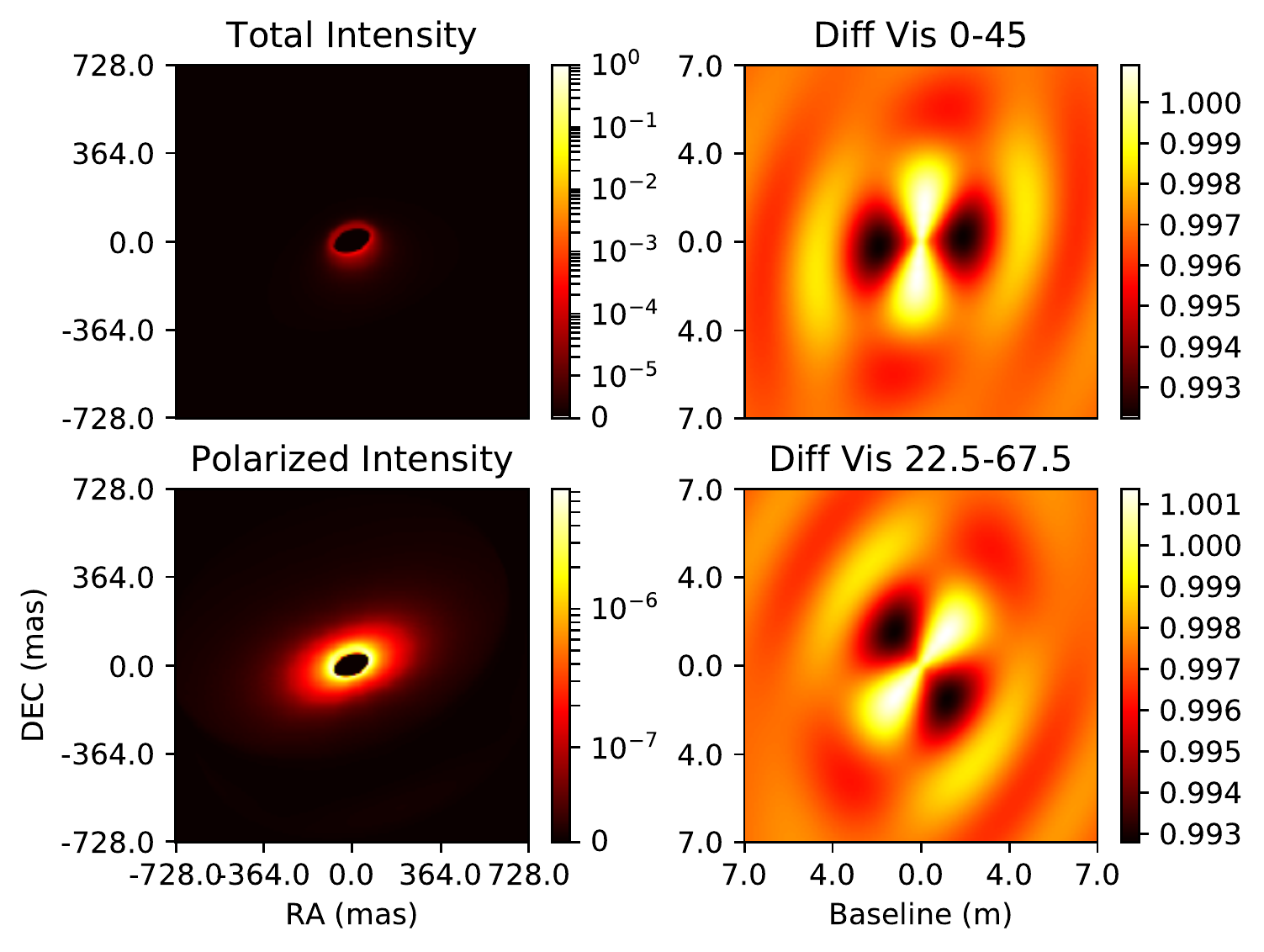}
    \includegraphics[width=3.4in]{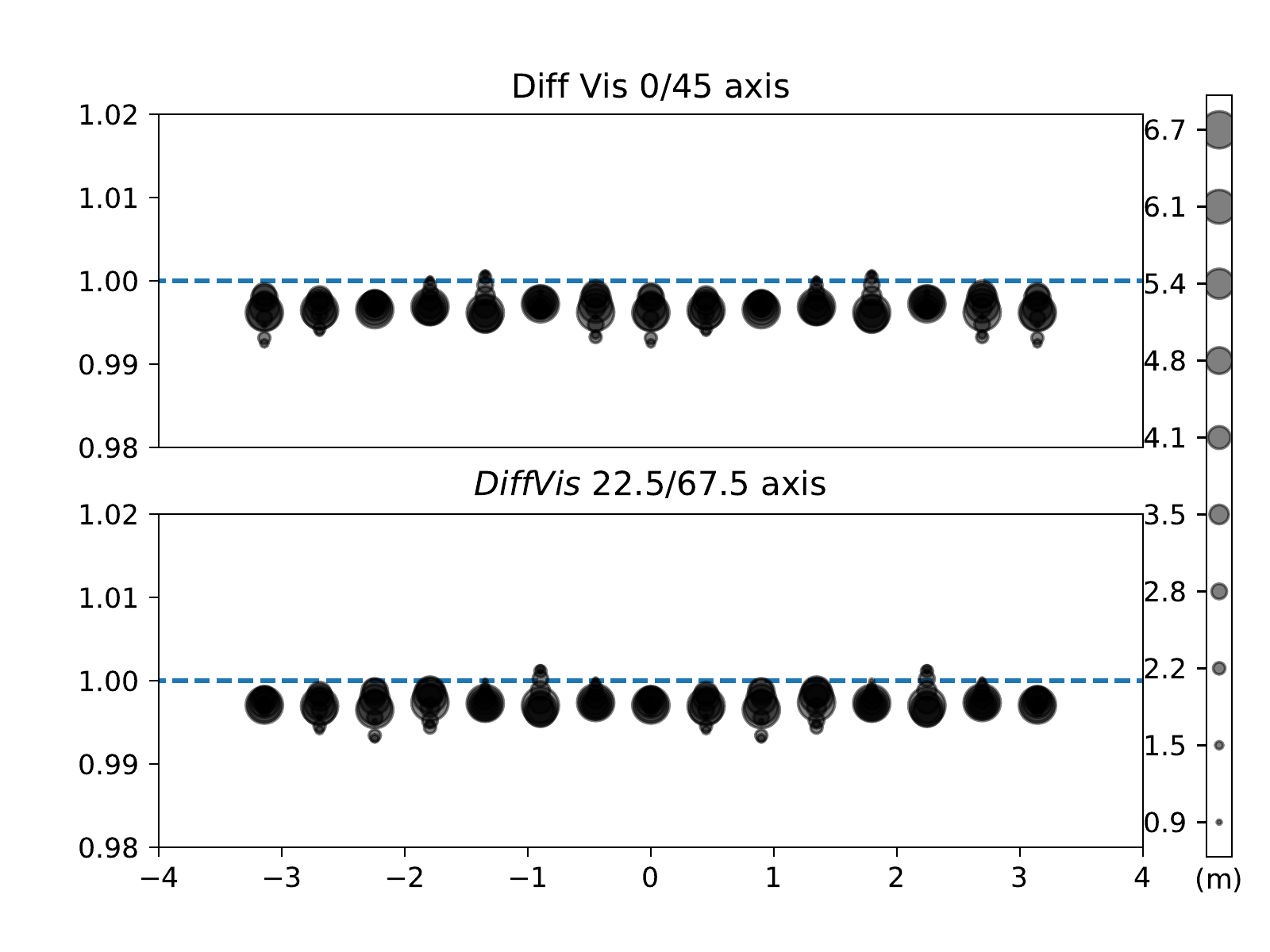}
    \caption{\textbf{Left-most column}: Disk model total and polarized
intensity normalized to the same value. \textbf{Middle column:}~Differential
visibilities shown in spatial frequency space. \textbf{Right}: Continuously
sampling the differential visibilities in azimuth. The NRM more sparsely
samples this space.}
    \label{fig:mcfost}
\end{figure}

To provide context for our reported precision we compared simulated data of a
disk generated with MCFOST \cite{pinte2006, pinte2009} based on modeling the
features in HD 97048~\cite{lagage2006, doucet2007}. For the purpose of this
simulation we place the inner disk edge at $\sim80~$mas extending out to
$\sim700~$mas. At a distance this corresponds to an inner edge at $\sim9~$au,
extending out to $\sim116~$au.  Figure \ref{fig:mcfost} shows the model of the
disk in total and polarized intensity as well as the ``perfect" differential
visibilities (Equation \ref{eqn:diffang}) over continuous spatial frequencies,
by taking the Fourier transformation of the simulated disk Stokes parameters.
The greatest azimuthal variation occurs between baselines of 1-2 m, where the
disk is the most resolved. Given the symmetry of the disk, the differential
closure phase signal is small ($<1^\circ$). 

To interpret differential visibility data it is helpful to forward model the
resolved polarized structure and compare this to differential visbilities
measured on sky. We outline  the steps to generate a simulated set of GPI NRM
data, converting from given Stokes I-V parameters, to linear polarization
images at four half-wave plate angles $0^o$, $22.5^o$, $45^o$, and $67.5^o$.
The intensity images are computed as follows:
\begin{eqnarray}
I_\pm(0^o) = \frac{I\pm Q}{2} ; 
I_\pm(22.5^o) = \frac{I\pm U}{2} \nonumber \\
I_\pm(45^o) = \frac{I\mp Q}{2} ; 
I_\pm(67.5^o) = \frac{I\mp U}{2} \nonumber \\
\end{eqnarray}
where (+,-) denotes the two polarization channels split by the Wollaston prism.
Each intensity is convolved with the GPI PSF accounting for some photon noise
and a small amount of jitter by convolving the image with a Gaussian of the
size of the desired jitter. In practice, however, vibrations are not uniform in
time or direction. They are mitigated by observing in low-wind conditions. 

\begin{figure}
    \centering
    \includegraphics[width=3.4in]{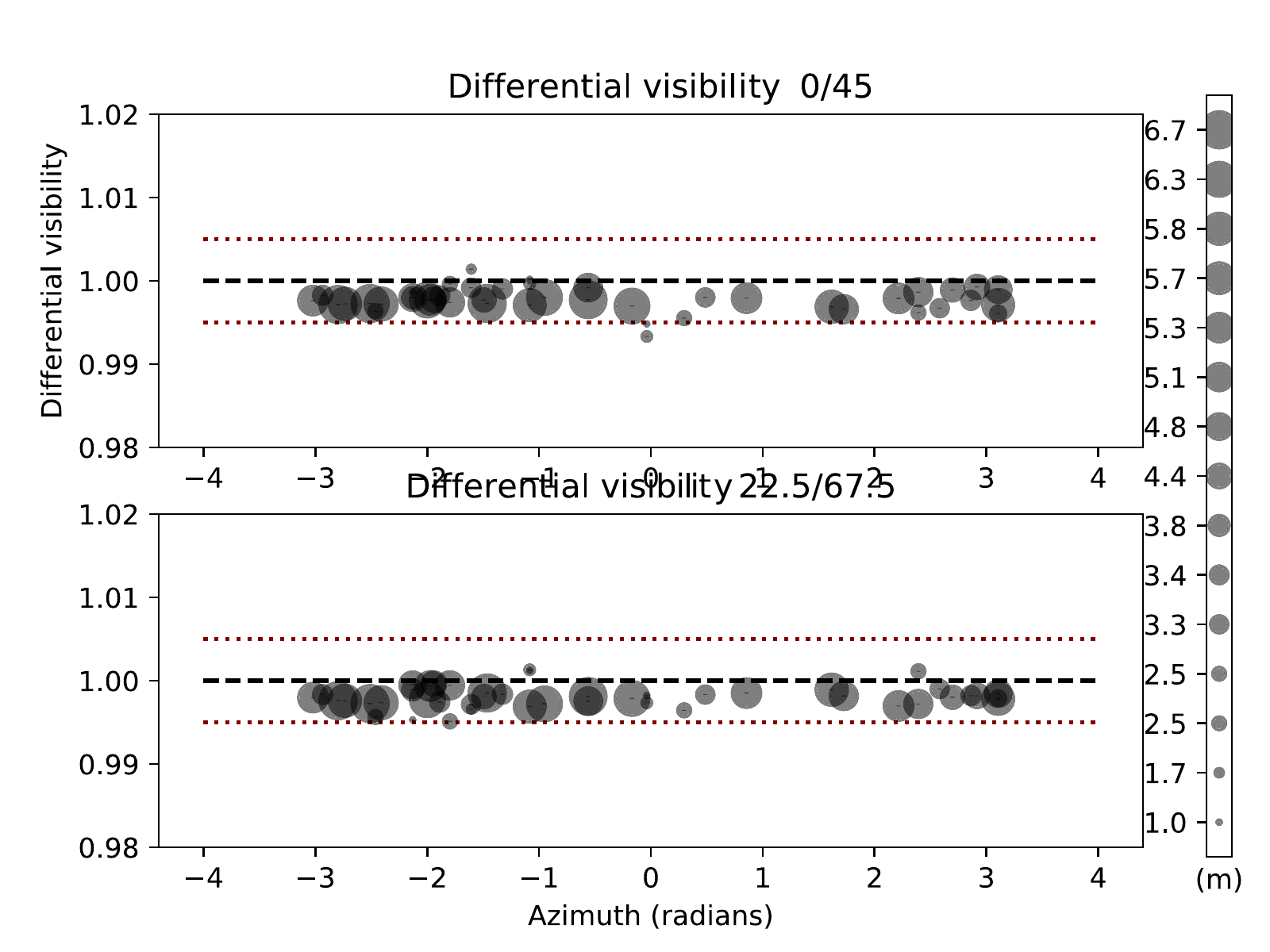}
    \includegraphics[width=3.4in]{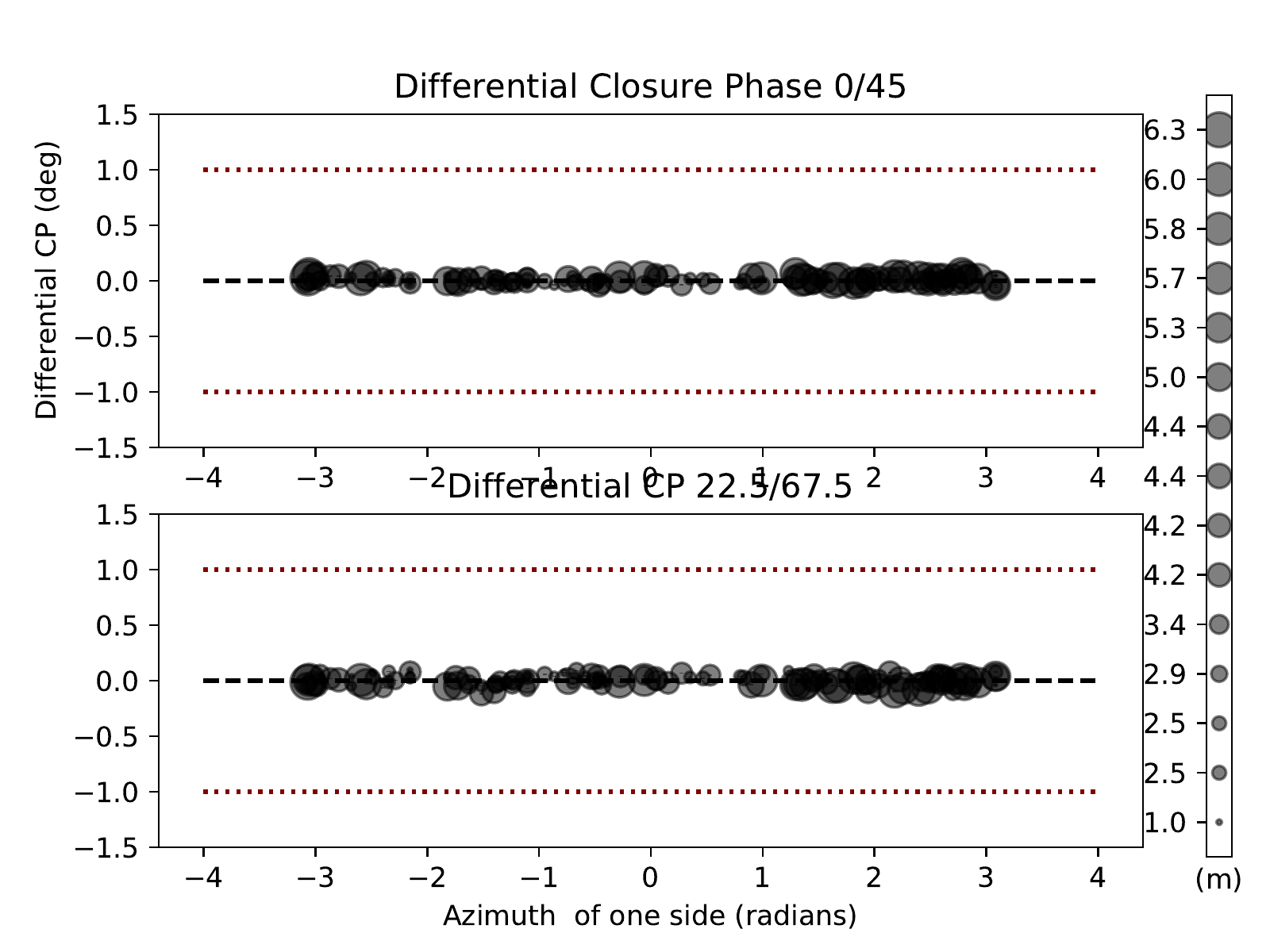}
    \caption{Differential observables measured from simulated low-noise GPI
data containing the disk model. Marker sizes denote the baseline length, where
the largest circles represent the longest baseline formed by the mask.
\textbf{The left} shows the measured differential visibilities. The red dotted
lines mark 0.5\% scatter, such as differential visibilities measured in the
best case for an unresolved star. \textbf{The right} shows differential closure
phases measured from the same simulated data. The red dotted lines mark
$1^\circ$ of scatter, similar to the scatter shown in Figure
\ref{fig:polperformance}. For polarized structure that is highly symmetric,
differential visibilities are a more sensitive observable. } 
    \label{fig:simulated_diffvis}
\end{figure}

We measure fringe observables from the simulated data and compare these to the
``perfect" visibilities in Figure \ref{fig:simulated_diffvis}. In this low-noise
simulation we can clearly see the disk and inner cavity are resolved by the raw
visibilities. This will not necessarily be the sky with real data, especially
when large vibrations are present. The mask baselines show similar variation
with baseline orientation as the ``perfect" visibilites. In practice,
reconstructing disk features will likely rely on forward modeling of the disk
as we show here~\citep[First described in][]{norris2012spie, norris2015}.

\bibliography{ms}

\clearpage

\end{document}